\documentclass[letterpaper]{article} % DO NOT CHANGE THIS
\usepackage{fullpage}
\usepackage{times}  % DO NOT CHANGE THIS
\usepackage{helvet}  % DO NOT CHANGE THIS
\usepackage{courier}  % DO NOT CHANGE THIS
\usepackage[hyphens]{url}  % DO NOT CHANGE THIS
\usepackage{graphicx} % DO NOT CHANGE THIS
\usepackage{natbib}  % DO NOT CHANGE THIS AND DO NOT ADD ANY OPTIONS TO IT
\usepackage{caption} % DO NOT CHANGE THIS AND DO NOT ADD ANY OPTIONS TO IT
\usepackage{hyperref}

\usepackage{algorithm}
\usepackage{algorithmic}
\usepackage{dsfont}

\usepackage{newfloat}
\usepackage{listings}
\DeclareCaptionStyle{ruled}{labelfont=normalfont,labelsep=colon,strut=off} % DO NOT CHANGE THIS
\floatstyle{ruled}
\newfloat{listing}{tb}{lst}{}
\floatname{listing}{Listing}

\usepackage{bibentry}

\usepackage[utf8]{inputenc} % allow utf-8 input
\usepackage{booktabs}       % professional-quality tables
\usepackage{amsfonts}       % blackboard math symbols
\usepackage{nicefrac}       % compact symbols for 1/2, etc.
\usepackage{microtype}      % microtypography
\usepackage{enumitem}
\usepackage{amsmath}
\usepackage{amssymb}
\usepackage{amsthm}
\usepackage{multirow}
\usepackage{mathrsfs}
\usepackage[dvipsnames]{xcolor}
\usepackage{mathtools, nccmath}
\usepackage{eqparbox}
\usepackage{comment}
\usepackage{subcaption}
\usepackage{cleveref}

\usepackage{pgfplots}

\usepackage{xspace}

\newcommand{\llm}{\textsf{GenAI}\xspace}
\newcommand{\poa}{\textnormal{PoA}\xspace}
\newcommand{\forum}{\textsf{Forum}\xspace}

\newcommand{\rg}{R^c\xspace}
\newcommand{\ro}{R^s\xspace}
\newcommand{\rvg}{r\xspace}

\DeclareMathOperator*{\argmax}{arg\,max}

\newcommand{\bm}[1]{\mathbf{#1}}

\newcommand{\appnx}[1]{{\ifnum\Includeappendix=1{#1}\else{the appendix}\fi}}

\makeatletter
\newcommand{\tpmod}[1]{{\@displayfalse\pmod{#1}}}
\makeatother

\newtheorem{theorem}{Theorem}
\newtheorem{lemma}{Lemma}

\newtheorem{definition}{Definition}
\newtheorem{proposition}{Proposition}

\newtheorem{corollary}{Corollary}
\newtheorem{example}{Example}

\newtheorem{observation}{Observation}

\DeclarePairedDelimiter\abs{\lvert}{\rvert}%
\newcommand{\tupbracket}[1]{\left\langle {#1} \right\rangle}

\newenvironment{proofof}[1]{\begin{proof}[\textnormal{\textbf{Proof of \Cref{#1}}}]}{\end{proof}} 

\newenvironment{sketch}[1]{\begin{proof}[\textnormal{\textbf{Proof sketch of \Cref{#1}}}]}{\end{proof}} 

\newcommand\ceil[1]{\left\lceil#1\right\rceil}
\newcommand\floor[1]{\left\lfloor#1\right\rfloor}
\newcommand{\ind}{\mathds{1}}

\newcommand{\ovu}{\overline{u}}
\newcommand{\unu}{\underline{u}}
\newcommand{\ovq}{\overline{q}}
\newcommand{\unq}{\underline{q}}

\newcount\Includeappendix 
\title{Braess's Paradox of Generative AI}
\author{
Boaz Taitler%
\thanks{%
    {Technion---Israel Institute of Technology (\url{boaztaitler@campus.technion.ac.il})}}
\and Omer Ben{-}Porat%
\thanks{%
    {Technion---Israel Institute of Technology (\url{omerbp@technion.ac.il})}}
}

\begin{document}

\maketitle

\sloppy

\begin{abstract}
ChatGPT has established Generative AI (GenAI) as a significant technological advancement. However, GenAI's intricate relationship with competing platforms and its downstream impact on users remains under-explored. This paper initiates the study of GenAI's long-term social impact resulting from the weakening network effect of human-based platforms like Stack Overflow. First, we study GenAI's revenue-maximization optimization problem. We develop an approximately optimal solution and show that the optimal solution has a non-cyclic structure. Then, we analyze the social impact, showing that GenAI could be socially harmful. Specifically, we present an analog to Braess's paradox in which all users would be better off without GenAI. Finally, we develop necessary and sufficient conditions for a regulator with incomplete information to ensure that GenAI is socially beneficial.
\end{abstract}

\section{Introduction}
ChatGPT has made Generative AI (GenAI) a household name, capturing the attention of the general public as the first fully operational GenAI~\citep{naveed2023comprehensive, chang2023survey}. Its widespread adoption stems from its capability to generate coherent texts and answer complex queries, showcasing human-like performance and even surpassing it~\citep{katz2024gpt}. Beyond mere functionality, GenAI has revolutionized various applications, ranging from automating customer service interactions to facilitating creative writing endeavors. This transformative impact is underpinned by GenAI's strengths, which are attributed to an extensive corpus used during its training phase, encompassing many fields and languages.

In contrast to those benefits, there is a growing body of work raising various concerns regarding GenAI, such as biases~\citep{abid2021persistent}, abusive and inappropriate content~\citep{solaiman2023evaluating}, and negative effects on users' behavior~\citep{dvorak2024generative, abbas2024harmful}. On top of those concerns, frequent training on high-quality data is a major issue~\citep{chen2023chatgpt}. Updated training data is crucial to the quality of GenAI's answers, necessitating the continual incorporation of new data to reflect current trends and changes.

To illustrate, consider ChatGPT and Stack Overflow as competing platforms. ChatGPT utilizes the questions and answers on Stack Overflow as its training data, providing users with a quick answer tailored to their questions. Consequently, users shift towards ChatGPT and neglect Stack Overflow as a forum for their queries (as witnessed by recent work~\citep{del2023large}). Over time, without additional training, the absence of fresh training data hampers ChatGPT's ability to provide accurate responses about new programming packages, while the diminished user activity on Stack Overflow affects both the volume and quality of answers available on the platform. This harmful combination negatively impacts the welfare of users, who suffer from outdated information and a lack of engagement in previously thriving community forums.

While the literature on ChatGPT and, more broadly, GenAI grows at an exponentially rapid pace, most of the research on ChatGPT since its release has focused on its performance~\citep{frieder2024mathematical, kocon2023chatgpt, chen2023chatgpt} and its applications across diverse domains \citep{kasneci2023chatgpt, liu2024your}. However, research has allocated scant attention, if any, to its social impacts concerning its intricate relationship with other competing platforms. Although GenAI is unanimously useful, its social impact prompts a critical research question:  Is the existence of GenAI as Q\&A platforms genuinely beneficial to its users, or would they be better off without it? Specifically, is the way we use GenAI socially beneficial in the long term?

\paragraph{Our contribution}
This paper initiates the study on GenAI's long-term social impact due to the weakening network effect of human-based platforms like Stack Overflow. We propose a sequential model that includes two platforms: A generative AI-powered service we term \llm, and a human-based forum called \forum in which users can ask questions and answer questions of others. In our model, \llm gains revenue from increased user traffic and incurs training and maintenance costs. \llm chooses a \emph{training scheme}: Strategically deciding in which rounds to train on fresh data. In contrast, \forum is a passive (i.e., non-strategic). A population of users utilizes \llm and \forum. We assume that user utility from \llm decreases between training rounds. We further assume that \forum has a network effect: User utility increases as the proportion of users using it rises. In our model, users choose according to a softmax function of the utilities from \llm and \forum, allowing users to have different sensitivity levels to their utility.

First, we address \llm's revenue-maximization task. We provide an efficient approximately optimal algorithm and show that the optimal training strategy has an unstable structure. Secondly, we examine the social impact. We say that a training scheme is \emph{socially harmful} if users are better off in a world without \llm, i.e., if users use \forum solely. We show the following surprising phenomena. 
\begin{theorem}[Braess's paradox of generative AI]\label{our cont thm possible harmful}
The optimal training scheme of \llm could be socially harmful. 
\end{theorem} 
Our result resembles Braess's paradox~\citep{braess1968paradoxon}: A paradoxical behavior of transport networks, where adding an extra road can increase travel time. Analogously, even though generative AI provides high-quality service that beats human-based alternatives in the short term, it could lead to deteriorated welfare in the long term. This counter-intuitive observation stems from \llm's negative impact on the network effect of \forum. After enough time without training, users who use \llm receive low-quality service but are locked into \llm, since \forum provides low-quality service as well without a strong user traffic. Since \llm wishes to maximize its revenue, it could train sparsely and still gain most of the user traffic. We further analyze the Price of Anarchy~\citep{roughgarden2005selfish,koutsoupias1999worst} and show that it is unbounded.

Thirdly, we adopt the point of view of a regulator. Theorem~\ref{our cont thm possible harmful} implies that regulation could be required to ensure that social welfare is greater than the counterfactual welfare in a world without \llm. Since the regulator is typically not informed about \llm's training scheme, ensuring \llm is socially beneficial poses a challenge. To address this, we develop necessary and sufficient conditions for social benefit and show that they are tight.

\subsection{Related Work}\label{subsec:related}
Our work is inspired by the mass adoption of generative AI, contributing to an emergent line of work on foundation models and game theory~\citep{laufer2024fine,yao2024human,conitzerposition,dean2024accounting}. This line of work includes challenges in the training process inspired by social choice theory~\citep{conitzerposition} and mechanism design~\citep{sun2024mechanism}, ways to cut training costs by pooling~\citep{huang2023train}, and revenue sharing between different actors~\citep{laufer2024fine}. In all of these cases, as argued by \citet{dean2024accounting}, planners should aim to understand societal impacts via mathematical modeling. Most relevant to our work is a recent work on content creation competition~\citep{yao2024human}. It models a Tullock contest~\citep{Tullock1980} between content creators, where some content creators use generative AI to create high-quality content, competing for users' engagement. The quality of the AI-generated content depends on the quality of the human-generated content. In a broader perspective, our work relates to social considerations in Machine Learning (see, e.g., \citep{caton2024fairness} for a recent survey). In this literature, social planners aim to restrict the output of a machine learning-based system to achieve long-term social welfare.

Our work also considers competition between platforms, a well-established topic in the economic literature~\citep{rietveld2021platform,karle2020segmentation,bergemann2024data}. In the computer science literature, recent works study the equilibrium structure in competition between strategic cloud providers~\citep{ashlagi2010competing} and optimization algorithms~\citep{immorlica2011dueling}. Other work studies competition in multi-learner settings~ \citep{dean2024emergent,ginart2021competing,ben2019regression,aridor2019perils}. Additionally, a recent work analyzes how social welfare behaves in data-driven marketplaces~\citep{jagadeesan2023competition}, which is similar to our view of \llm. 

Finally, a crucial part of our model is \forum's network effect, which is inspired by the vast economic literature on this topic~\citep{katz1985network, rochet2003platform,katz1986technology,feldman2013competition}.

%%%%%%%%%%%%%%%%%%%%%%%%%%%%%%%%%%%%%%%%%%%%%%%%%%%%%%%%%%%%%%%%%%%%%%%
%
%
%                         Model
%
%
%%%%%%%%%%%%%%%%%%%%%%%%%%%%%%%%%%%%%%%%%%%%%%%%%%%%%%%%%%%%%%%%%%%%%%%

\section{Model} \label{model section}

\paragraph{General overview of the ecosystem} The setting evolves over $T$ discrete rounds, where in each round, a population of users chooses between \llm and \forum for posing their questions. \llm and \forum 
represent a Generative AI system and a Q\&A forum, respectively. We take the perspective of \llm, whose decisions are training times. In each round $t$, we denote by $x_t \in \{0, 1\}$ whether \llm trains or not. We further let $\bm x = (x_1, \dots x_T)$ denote the \emph{training scheme} of \llm and always assume that \llm trains on the first round, that is, $x_1 = 1$. Users' decisions are stochastic; we let $p_t$ denote the proportion of users selecting \llm in round $t$, with the complementary $1-p_t$ selecting \forum. As we explain later, $p_t$ depends on several elements including the training scheme $\bm x$, i.e., $p_t=p_t(\bm x)$.  

An instance of our problem is represented by the tuple $\tupbracket{\rvg, c_m, c_{train}, \rg,  \ro, \beta, p_1}$; we now elaborate on the components of the model.

\paragraph{\llm}
The \emph{revenue} of \llm consists of three key components. Firstly, \llm gets a reward of $\rvg \in \mathbb{R}_+$ from users utilizing it, where $r$ could represent direct payment, indirect income from advertising, etc. Secondly, \llm pays a maintenance cost $c_m \in \mathbb{R}_+$. Thirdly, every time \llm trains it incurs an additional cost of $c_{train}  \in \mathbb{R}_{+}$. Altogether, the revenue of \llm in round $t$ is $v_t(\bm x)$, where
\[
v_t(\bm x) = p_t(\bm x) \rvg - c_m - x_t c_{train}.
\]
We further let $V$ denote the total revenue over the $T$ rounds, namely, $V(\bm x) = \frac{1}{T} \sum_{t=1}^T v_t(\bm x)$. 

Users gain \emph{utility} from using \llm. We let $\rg : \mathbb{N} \rightarrow \mathbb{R}_{+}$ denote this utility, and assume that it decreases as the time since the last training increases. That is, $\rg(t - \tau)$ is the utility in round $t$ if the last training of \llm occurred in round~$\tau<t$, which is decreasing in $t$. The gradual utility decrease captures several phenomena, such as outdated data like the identity of current Olympic medalists, or new technological developments that did not exist in its training data, like newly developed Python packages. Indeed, this assumption is well-supported empirically~\citep{chen2023chatgpt}. Since user interaction with $\llm$  involves queries that yield varying degrees of relevance and timeliness, we think of $\rg$ as the average utility users gain from \llm. Other than being monotonically decreasing with respect to the last training time, we have no assumptions on the structure of $\rg$.

As we mentioned before, \llm strategically picks a training scheme $\bm x$. For notational convenience, we define $\mathcal{T}(\bm x) = \{t | x_t = 1, t\in [T]\}$ to be the set of all training rounds in $\bm x$. Further, we let $\gamma_t=\gamma_t(\bm x)$ denote the time between round $t$ and its last training before that round, i.e., $\gamma_t(\bm x) = t - \max\left\{\mathcal{T}(\bm x)\cap [t]\right\}$. Therefore, we simplify our notations and use $\rg(\gamma_t(\bm x))$ to denote the user utility in round $t$.

\paragraph{\forum}
The revenue of \forum is also derived from the users who use it. We assume \forum is non-strategic, as its actions could be rather complex, and it is unclear whether or how such actions affect the ecosystem (see discussion in Section~\ref{sec:discussion}).\footnote{For instance,  \forum could sell data or develop its own GenAI. While working on this paper, it was reported that Stack Overflow partnered with OpenAI~\citep{stack+gpt}.} We denote by $\ro : [0, 1] \rightarrow \mathbb{R}_{+}$ the average user utility from \forum in round $t$. We assume that $\ro$ is time-invariant but is influenced by  \emph{network effects}, a well-established phenomenon empirically demonstrated in various studies~\citep{mcintyre2017networks, katona2011network}. That is, the utility $\ro$ increases with the proportion of users $1-p_t$ using \forum. 
To ease notation, we denote $\ro=\ro(p_t)$, which means $\ro$ is decreasing as more users use \llm. 
As before, $\ro(p_t)$ could have any general form, and we only assume it is a decreasing and differentiable function.

\paragraph{User behavior}
The \emph{instantaneous welfare} in round $t$ is the (weighted) average utility of users using \llm and \forum, which is given by
\begin{equation}\label{eq: u_t}
u_t(\bm x) = p_t(\bm x) \rg(\gamma_t(\bm x)) + (1-p_t(\bm x) ) \ro(p_t(\bm x)).   
\end{equation}
The (cumulative) \emph{social welfare} $U:\{0,1 \}^T \rightarrow \mathbb R_+$ is the sum of the instantaneous welfare over all rounds, $U(\bm x) = \sum_{t = 1}^{T} u_t(\bm x)$. 

We now describe the proportions $(p_t)_t$. For $t=1$, the proportion $p_1$ of users using \llm is a model parameter. We let this quantity be determined exogenously as it could incorporate users' willingness to be early adopters, reflecting the population's appetite for innovation and novelty. For $t >1$, we assume that user decisions are stochastic---they assign probabilities to \llm and \forum based on the utilities the platforms yield. Further, users are Markovian: They base their decisions on the rewards from the preceding round $t-1$, but are otherwise independent of their past decisions.  We formulate $p_t$ as a softmax function of the utilities from \llm and \forum from the last round, 
\begin{equation}\label{eq:markovian update}
p_t(\bm x) = \frac{\exp(\beta \rg(\gamma_{t-1}(\bm x)))}{\exp(\beta \rg(\gamma_{t-1}(\bm x))) + \exp(\beta \ro(p_{t-1}(\bm x)))},    
\end{equation}
where the temperature $\beta$ is the decision \emph{sensitivity} parameter, determining how sensitive users are to their utility. For instance, 
$\beta \rightarrow \infty$ suggests that users are utility maximizers, whereas $\beta=0$ indicates that users are indifferent to their utility and choose uniformly between \llm and \forum. 
\paragraph{Assumptions}
Recall that we assume that $\rg$ decreases with the time from the last training and that $\ro$ decreases as more users rely on \llm. Further, for our model to be realistic, we need to make sure that both \llm and \forum could be attractive to users. Formally, we assume that there exists $t\in \mathbb N$ such that 
\begin{equation}\label{eq:utility assumption}
\rg(t) < \ro(0) < \rg(0).   
\end{equation}

The left inequality implies that \llm becomes inferior if enough time has passed since its last training, while the right inequality suggests that \llm is better than \forum immediately after training.
\begin{example}\label{example}
\normalfont 
%$\tupbracket{\rvg, c_m, c_{train}, \rg,  \ro, \beta, p_0}$
Consider an instance with \llm parameters $\rvg = 1$, $c_m = 0.6$, $c_{train} = 0.504$, utility functions $\rg(t) = 3\cdot0.5^t$, $\ro(p) = 1-p$, and user behavior parameters $\beta = 1$, $p_1 = 1$. Let the horizon be $T = 20$, and consider the training scheme that only train at $t = 1$, i.e., $\bm x^0 = (x^0_1,\dots,x^0_T)= (1, 0, \ldots, 0)$. At $t = 1$, the proportion is $p_1(\bm x^0) = p_1$, and the number of rounds from the last training round is $\gamma_1 = 0$; therefore,
\begin{align*}
u_1(\bm x^0) &= p_1 \rg(0) + (1-p_1) \ro(p_1) = 1 \cdot 3\cdot0.5^0 = 3.
\end{align*}
Notice that if all the users were to use \forum, then the users' instantaneous welfare would have been $\ro(0) = 1$, which is lower than $u_1(\bm x^0)$. Thus, at $t = 1$, the presence of \llm is socially beneficial. Next,  \llm's revenue in round 1 is
\begin{align*}
v_1(\bm x^0) &= p_1 \rvg - c_m - x^0_1 c_{train}  = -0.104.
\end{align*}
Moving on to round $t=2$, we compute the proportion $p_2(\bm x^0)$. The proportion $p_2(\bm x^0)$ depends on $\rg(\gamma_1)=\rg(0)$ and $\ro(p_1(\bm x^0))=\ro(1)$; therefore,
\begin{align*}
p_2(\bm x^0) = \frac{e^{\beta \rg(0)}}{e^{\beta \rg(0)} + e^{\beta \ro(1)}} = 0.95.
\end{align*}
In the second round, $\gamma_2=1$ since $x^0_2=0$ and \llm does not train. Consequently, we have all the necessary information to compute $u_2(\bm x^0)$ and $v_2(\bm x^0)$.

we highlight two additional training schemes. We let $\bm x^r$ represent the revenue-maximizing scheme, which can be shown to train in rounds $t\in \{ 1,4,7,9,12,14,17 \}$. Additionally, let $\bm x^w$ represent the socially optimal scheme, training in every round. Figure~\ref{fig:example 1} demonstrates the social welfare of the three schemes over time, and the \emph{counterfactual} welfare $t\ro(0) $ obtained when users can only use \forum.

We observe several findings. First, the three schemes generate slightly higher welfare in the first round than the counterfactual due to our assumption in Inequality~\eqref{eq:utility assumption}. Second, as we expect, $\bm x^w$ generates the highest welfare over time. Third, $\bm x^0$ provides higher welfare than the counterfactual till $t=6$, and then they reverse. Namely, $\bm x^0$ makes users worse off. In that regard, notice that $\bm x^r$, the revenue-maximizing scheme, turns out to be better than the counterfactual in the long run.
\begin{figure}[t]
    \centering

\def\datagraphexample{figures/example.csv}

\definecolor{O2}{RGB}{0,0,255}
\def\unitwo{O2}

\definecolor{O1}{RGB}{255,0,0}
\def\unicolor{O1}

\definecolor{O3}{RGB}{0,0,0}
\def\unithree{O3}

\definecolor{O4}{RGB}{0,255,0}
\def\unifour{O4}

\begin{tikzpicture}[scale=0.85]

\begin{axis}[
    axis lines = left,
    xlabel = Round $t$,
    ylabel = Welfare $U(\bm x)$,
    legend pos=north west,
    legend cell align = left
]
\addplot [\unitwo, mark=*, mark options={solid}, line width=1.1pt] table[x=steps, y=U_nt, col sep=comma] {\datagraphexample};
\addlegendentry{$\bm x^0$}

\addplot [\unithree, mark=square*, mark options={solid}, line width=1.1pt] table[x=steps, y=U_rev, col sep=comma] {\datagraphexample};
\addlegendentry{$\bm x^r$}

\addplot [\unifour, mark=triangle*, mark options={solid}, line width=1.1pt] table[x=steps, y=U_uo, col sep=comma] {\datagraphexample};
\addlegendentry{$\bm x^w$}

\addplot [\unicolor, mark=diamond*, mark options={solid}, line width=1.1pt] table[x=steps, y=U_s, col sep=comma] {\datagraphexample};
\addlegendentry{counterfactual}

\end{axis}

\end{tikzpicture}
    \caption{Social welfare over time in Example~\ref{example}. The schemes $\bm x^0$, $\bm x^r$, and $\bm x^w$ are the no-training, revenue-maximizing, and welfare-maximizing schemes, respectively. The counterfactual welfare is the welfare users obtain in a world without \llm.}
    \label{fig:example 1}
\end{figure}
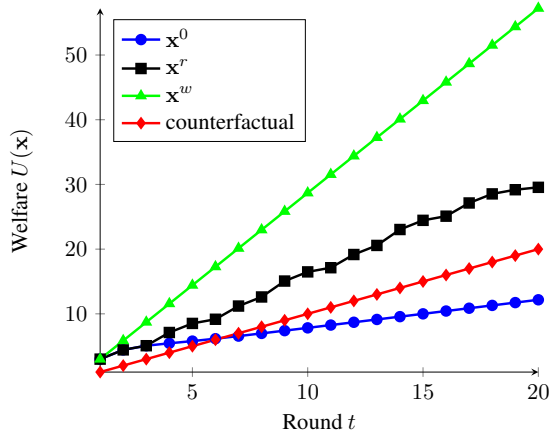

\end{example}

\section{\llm's Revenue Maximization and Cyclic Training Schemes} \label{llm opt section}
In this subsection, we consider \llm's revenue-maximization problem. We start by developing a dynamic programming-based approach to approximate the optimal revenue. Then, we highlight a class of simple training schemes---cyclic schemes, which train every fixed number of rounds. We show a surprising result: Cyclic schemes could be highly sub-optimal.

Recall that training schemes are binary vectors; hence, we can inefficiently find \llm's revenue-maximizing scheme by considering all $2^T$ options. However, as we show next, there is an efficient algorithm that approximates the optimal revenue for a broad class of instances. We defer the full description of the algorithm to \appnx{Appendix~\ref{llm opt section}}, and only write its formal guarantees.
\begin{proposition}\label{prop:dynamic}
Fix any instance such that $\ro$ is $L$-Lipschitz satisfying $\ro(1) = 0$ and $\beta \cdot L < \frac{16}{7}$. For any $\varepsilon>0$ such that $\varepsilon < \frac{16-7\beta L}{14 \beta L e^{\beta L} T}$, there exists an algorithm that runs in $O\left(\frac{T^2}{\varepsilon}\right)$ time and returns a scheme $\bm x$ satisfying $V(\bm x) \geq \max_{\bm x'} V(\bm x')-\varepsilon \rvg T$.
\end{proposition}
Despite that Proposition~\ref{prop:dynamic} provides a way to approximate \llm's revenue, such an approach has several weaknesses. First, from a practical standpoint, polynomial runtime can still be infeasible if the horizon $T$ is large. Second, solutions obtained by dynamic programming can be uninterpretable, possibly hiding issues like large deviations in welfare over time. And third, \llm's revenue may occasionally be negative for several consecutive rounds. Although we do not require a positive revenue balance over time in our model, this can be a weakness in real-world scenarios. 

To that end, we examine a wide class of \emph{cyclic training schemes}. As we show, cyclic training schemes offer certainty in both user utility and revenue. They create cycles in which utility and revenue can vary, but they become (asymptotically) constant over different cycles. Furthermore, cyclic training schemes could enjoy planned obsolescence \citep{bulow1986economic, lee1998theory}---the practice of releasing new product versions at fixed intervals to encourage repeated purchases and maintain market share. The formal definition of a cyclic training scheme is as follows.
\begin{definition}[$k-cyclic$ training scheme] \label{cyclic training scheme}
We say that $\bm x$ is \emph{$k$-cyclic} for $k \in \mathbb N$ if
\begin{align*}
x_t = \begin{cases}
    1 & \mbox{$t \tpmod{k} = 1$} \\
    0 & \mbox{otherwise}
\end{cases}.
\end{align*}
\end{definition}
Note that for all $k \in \mathbb N$, a $k$-cyclic scheme always trains in the first round $t=1$, which is consistent with our definition of a scheme.\footnote{In our formal statements, we address a more general definition of cyclic schemes, allowing the prefix and suffix of a cyclic scheme to behave arbitrarily. However, as long as their length is constant w.r.t. to $T$, all of our results still hold; thus, we focus on this simpler form here.} As the next proposition shows, cyclic training schemes are stable over time. 
\begin{proposition} \label{prop: proportions converge}
Fix an instance and $k \in \mathbb N$. There exists a sequence  of constants $(q^\star_i)_{i=1}^k$ such that for every $i \in [k]$, it holds that 
\[
\lim_{l,T\rightarrow \infty} p_{l\cdot k +i}(\bm x_T^k)= q_i^\star,
\]
where $\bm x_T^k$ is the $k$-cyclic scheme for $T$ rounds.
\end{proposition}
As an immediate corollary of this proposition, we get that
\begin{corollary}\label{cor:fixed revenue and welfare}
It holds that
\begin{itemize}
    \item $\lim_{l,T\rightarrow \infty} v_{l\cdot k +i}(\bm x_T^k)= q_i^\star \rvg - c_m - \ind_{i=1} c_{train}$
    \item $\lim_{l,T\rightarrow \infty} u_{l\cdot k +i}(\bm x_T^k) =  q_i^\star \rg(i-1) + (1-q_i^\star ) \ro(q_i^\star)$.
\end{itemize}
\end{corollary}
Corollary~\ref{cor:fixed revenue and welfare} ensures that both welfare and revenue are (asymptotically) constant in every step of every cycle. That is, they can vary inside a cycle but become constant in the same step $i$ across different cycles. 

Intuitively, one would expect cyclic schemes to be optimal, or at least approximately optimal. Indeed, optimizing the cycle length $k$ to maximize revenue-per-cycle seems to maximize the revenue $V$, as the revenue-per-cycle is guaranteed to converge. Surprisingly, this is not the case even for ``standard'' instances like the one in Example~\ref{example}.
\begin{theorem} \label{thm not cyclic}
Fix the instance in Example~\ref{example}. There exists $T^\star \in \mathbb{N}$, such that for any $T > T^\star$ and any cyclic training scheme $\bm x$,  it holds that
\[
\max_{\bm x' \in \{0,1,\}^T}V(\bm x') - V(\bm x) = \Theta(1).
\]
\end{theorem}
\iffalse
Before we provide a proof sketch, we explain the intuition behind it. Consider a $k$-cyclic and $k'$-cyclic schemes, $\bm x^k$ and $\bm x^{k'}$, respectively, for $k'>k$. Further, consider a non-cyclic scheme $\bm x^{k,k'}$ that alternates between training every $k$ and $k'$ rounds. Namely,
\begin{align*}
x^{k,k'} = \begin{cases}
    1 & \mbox{$t \tpmod{k+k'} \in \{1,k+1\}$} \\
    0 & \mbox{otherwise}
\end{cases}.
\end{align*}
At a first glance, it seems that the revenue of $\bm x^{k,k'}$ is a weighted average of the revenues of $\bm x^{k}$ and $\bm x^{k}$. Indeed, the costs of $\bm x^{k,k'}$ are a weighted average of the costs of $\bm x^{k}$ and $\bm x^{k}$. To see this, let $C(\bm x^k)$ be the costs of $\bm x^k$, i.e., $C(\bm x^k)=c_{train}\cdot \nicefrac{T}{K}+c_m T$. Similarly, let let $C(\bm x^{k'})$ be the costs of $\bm x^{k'}$. Then, the costs of $\bm x^{k,k'}$ can be written as
\begin{equation*}
C(\bm x^{k,k'}) = \frac{k}{k+k'} C(\bm x^k) + \frac{k'}{k+k'} C(\bm x^{k'}).
\end{equation*}
However, the rewards are non-linear. Specifically, the reward $\sum_{t=1}^T p_t(\bm x^{k,k'})$ could be higher or lower than the weighted average  
\[
\frac{k}{k+k'} \sum_{t=1}^T p_t(\bm x^{k}) + \frac{k'}{k+k'}\sum_{t=1}^T p_t(\bm x^{k'}).
\]
This occurs due to the non-linearity of the Markovian update of proportions (recall Equation~\eqref{eq:markovian update}).
\fi

\begin{sketch}{thm not cyclic}
The proof of the theorem is rather technical. To prove the theorem, we need to show that a non-cyclic scheme outperforms \emph{all} cyclic schemes by a non-negligible quantity. First, we present several key lemmas that allow us to focus on cycles with bounded length. Then, we analyze the convergence rate of the proportions from Proposition~\ref{prop: proportions converge} using \emph{contraction} of local neighborhoods. This allows us to upper-bound the revenue-per-cycle for any cycle length and thus the revenue. Finally, we provide another (non-cyclic) training scheme and use similar techniques to lower-bound its revenue. The proof is completed by showing that the lower bound of our non-cyclic scheme is better than the highest upper bound of cyclic schemes.
\end{sketch}

\section{Social Impact} \label{section social impact}
In this section, we explore societal implications, demonstrating that the presence of \llm can be \emph{socially harmful}, reducing social welfare compared to its absence. We begin with a formal definition of social harmfulness and then show a counter-intuitive phenomenon: Optimal \llm policies could be socially harmful. We flash out the main issue behind social harmfulness, which is prolonged periods without training. Later, Subsection~\ref{subsec: social PoA} analyzes the Price of Anarchy, quantifying the inefficiency due to strategic behavior.

Next, we formally define the notion of social benefit. 
\begin{definition}[Socially beneficial scheme] \label{def socially harmful}
We say that a training scheme $\bm x$ is \emph{socially beneficial} if 
\begin{equation}\label{eq:def harmful}
U(\bm x)  \geq  T\cdot \ro(0).    
\end{equation}
\end{definition}
Definition~\ref{def socially harmful} compares the social welfare with and without the presence of \llm. The left-hand side of Inequality~\eqref{eq:def harmful} is the social welfare in the presence of \llm under the scheme $\bm x$. The right-hand side represents the counterfactual social welfare in the absence of \llm, where all users choose \forum, similarly to the counterfactual welfare in Example~\ref{example}. Conversely, if Inequality~\eqref{eq:def harmful} does not hold, we say that $\bm x$ is \emph{socially harmful}. Interestingly, 
\begin{observation}[Braess's paradox of generative AI]\label{obs possible harmful}
There exist instances where the optimal training scheme $\bm x$ is socially harmful. 
\end{observation} 
Observation~\ref{obs possible harmful} suggests that using \llm can lead to deteriorated welfare, a paradox analogous to Braess's paradox in traffic networks~\cite {braess1968paradoxon}. Indeed, since \llm can potentially provide a better utility than \forum, why does it occur?

Two simultaneous forces cause this paradox. First, when users rely on \llm, they decrease their usage of \forum. Since user utility from \forum depends on network effects, lower usage results in reduced utility. And this is where the second force enters: To keep its dominance, \llm need not train much. After prolonged periods without training, its utility diminishes but is still better than the low utility from \forum since its network effect is weak. 

\subsection{Price of Anarchy} \label{subsec: social PoA}
In this subsection, we explore another useful way to quantify social harmfulness. We adopt the Price of Anarchy (\poa), measuring the inefficiency due to strategic behavior~\citep{koutsoupias1999worst,roughgarden2005selfish}. In our context, let $\mathcal X$ be the set of revenue-maximizing training schemes. For any instance $I$, the \poa is defined as
\begin{align*}
\poa(I) = \frac{\max_{\bm x} U(\bm x)}{\min_{\bm x \in \mathcal X}U(\bm x)}.
\end{align*}
The next proposition demonstrates that the \poa is unbounded.
\begin{proposition} \label{infinite poa}
For every $M \in \mathbb R_+$, there exists an instance $I$ with $\poa(I) > M$.    
\end{proposition}
\begin{sketch}{infinite poa}
To prove the proposition, we focus on the terms $\ro(1)$ and $\rg(t)$ for large values of $t$. The former is the utility from \forum when all users use \llm, while the latter is the utility from \llm after $t$ rounds without training. Consider an instance with parameters that satisfy $\rg(t) > 0$, $\ro(1) = 0$ and $\beta = \infty$. For this selection of sensitivity, the users are strategic, i.e., $p_{t}(\bm x) = 1$ if $\rg(\gamma_{t-1}) > \ro(p_{t-1}(\bm x))$. Due to our assumption in Equation~\eqref{eq:utility assumption}, this property pushes users to choose \llm with $p_t(\bm x) = 1$ for every $t$ and every training scheme as $\rg(t) > \ro(1) = 0$. \llm has no incentive to train, and therefore the no-training scheme $\bm x^0$ is the revenue-maximizing scheme, inducing $U(\bm x^0) = \sum_{t = 1}^T \rg(t-1)$. On the other hand, training in every round maximizes the welfare; hence, $\max_{\bm x}U(\bm x) = T\rg(0)$. Overall,  $\poa(I) = \frac{T\rg(0)}{\sum_{t = 1}^T \rg(t-1)}$, which could be arbitrarily large if $\lim_{t \rightarrow \infty}\rg(t) = 0$.

\end{sketch}

\iffalse
Therefore, to analyze social harmfulness, it is beneficial to observe the social welfare between training periods. Formally, let $U(\bm x, t, t')$ be the cumulative welfare between rounds $t$ and $t'$ for $t< t'$, i.e., $U(\bm x, t, t')= \sum_{i = t}^{t'-1} u_i(\bm x)$. Extending Definition~\ref{def socially harmful}, we say that $\bm x$ is \emph{$(t, t')$-socially harmful} if $U(\bm x, t, t') < (\tau'- \tau)\cdot \ro(0) $. 

Notice that a $(0,T)$-socially harmful scheme 

The below proposition shows that social harmfulness can occur in all instances, quantifying the training frequency that leads to harmfulness.  
\begin{proposition} \label{thm stop training}
For any instance, there exists a natural number $k \in \mathbb{N}$ that depends on the instance parameters such that all training schemes that train on two consecutive rounds $\tau,\tau' \in [T]$ with $\abs{\tau'-\tau}> k$ are $(\tau, \tau')$-socially harmful.
\end{proposition}
\fi

\section{Regulating Training Frequency}\label{subsec:regulator}
Despite that \llm could be socially harmful, as the previous section demonstrates, there are socially beneficial training schemes (e.g., training every round). 
In this section, we aim to identify interventions to \llm's scheme that could mitigate its harmful effects. We develop a set of requirements that could be imposed on \llm by a regulator.

The regulator's ability to intervene depends on its knowledge of the instance. We assume that the regulator has complete information on user-related parameters: The utility functions $\ro,\rg$ and sensitivity $\beta$. If the regulator could demand \llm to commit to a publicly known training scheme, it could verify that the scheme is socially beneficial. However, the regulator does not have access to any proprietary information belonging to \llm and \forum. In particular, the regulator lacks knowledge of the scheme $\bm x$ and the resulting proportions $(p_t)_t$.
\subsection{Welfare Between Two Consecutive Training Rounds}
Our approach to ensure social benefit is to bound the maximal number of rounds between training. In this subsection, we explain why such a regulation is fundamental in guaranteeing social benefit. Let $\tau_1,\dots, \tau_{\abs{\mathcal{T}}}$ be the ordered elements of the training rounds $\mathcal{T}$, and let $\tau_{\abs{\mathcal{T}}+1}=T+1$. A sufficient condition for social benefit is that the inequality
\begin{equation}\label{ineq: sufficient}
\sum_{t=\tau_i}^{\tau_{i+1}-1} u_t (\bm x)  \geq (\tau_{i+1}-\tau_i)\cdot \ro(0)
\end{equation}
would hold for every $i \in \mathcal{T}$. To see why, fix a scheme $\bm x$. Notice that if Inequality~\eqref{ineq: sufficient} holds, then
{
\thickmuskip=2mu plus 2mu
\[
U(\bm x) = \sum_{i=1}^{\mathcal{T}}\sum_{t=\tau_i}^{\tau_{i+1}-1} u_t (\bm x) \geq \sum_{i=1}^{\mathcal{T}}(\tau_{i+1}-\tau_i)\cdot \ro(0) = T \cdot \ro(0);
\]
}%
hence, $\bm x$ is socially beneficial by definition. As a result, ensuring Inequality~\eqref{ineq: sufficient} holds allows the regulator to ensure social benefit.

To that end, fix an arbitrary training round $\tau$ and let $\Delta$ denote the number of rounds between $\tau$ and the next training round (i.e., $\tau+\Delta \in \mathcal T$). Recall the definition of $u_t(\bm x)$ in Equation~\eqref{eq: u_t}. If the regulator knows $p_\tau(\bm x)$, it could compute $u_\tau(\bm x)$ as it only depends on $p_\tau(\bm x)$ and the fact that $\gamma_\tau=0$.\footnote{
To verify such a requirement is fulfilled, the regulator must be informed when \llm trains. This is a realistic assumption that is also studied in recent research~\citep{chen2023chatgpt}.} 

The regulator could also compute $p_{\tau+t}(\bm x)$ for any $t\in \{0,\dots, \Delta-1  \}$, due to the Markovian nature of the proportions; thus, it could compute $u_{\tau+t}(\bm x)$ for those rounds as well. To sum, if the regulator knows $p_\tau(\bm x)$, it could compute $\sum_{t=\tau}^{\tau+\Delta-1} u_t (\bm x)$ and determine whether Inequality~\eqref{ineq: sufficient} holds. However, as noted before, the regulator cannot generally access the proportions $(p_t)_t$. 

\subsection{Bounding the Proportions} \label{subsec: bounding proportions}
In this section, we develop non-trivial bounds on $\sum_{t=0}^{\Delta-1} u_{\tau+t} (\bm x)$. Specifically, we use crude bounds on $p_\tau(\bm x)$ for any scheme $\bm x$, and unravel the structure of the Markovian update of proportions to non-trivially bound $p_{\tau+t}(\bm x)$ for any $t\in \{0,\dots, \Delta-1 \}$. The next technical lemma demonstrates the monotonicity of the induced proportions. 
\begin{lemma}[Monotonicity]\label{prop greater p0}
Fix an instance and a training scheme $\bm x$, and let $(p_t)_{t=1}^T$ be the induced proportions. Further, let $(p_t')_{t=1}^T$ be the induced proportion of the same instance except that the initial proportion is $p'_1 > p_1$. Then, for every $t\in [T]$ we have $p'_t>p_t$.
\end{lemma}
Next, we introduce the auxiliary sequence $(q_t^\alpha)_{t=0}^{\infty}$ for every $\alpha \in [0,1]$. We define $q^\alpha_t$ such that
\[
q^\alpha_t = 
\begin{cases}
    \alpha & t=0\\
 \frac{e^{\beta \rg(t-1)}}{e^{\beta \rg(t-1)} + e^{\beta \ro(q^\alpha_{t-1})}} & t > 0   
\end{cases}.
\]
Note the resemblance between the recursive update of $q$ and the definition of $p_t$ in Equation~\eqref{eq:markovian update}. For instance, if $\alpha=p_\tau(\bm x)$, we get $q^\alpha_t = p_{\tau+t}(\bm x)$ for every $t\in \{0, \dots \Delta -1 \}$. 

\begin{figure*}[t]
\centering
\begin{subfigure}[t]{0.49\textwidth}
\centering

\def\datagraphpvst{figures/pvst.csv}

\definecolor{O2}{RGB}{0,0,255}
\def\unitwo{O2}

\definecolor{O1}{RGB}{255,0,0}
\def\unicolor{O1}

\definecolor{O3}{RGB}{0,0,0}
\def\unithree{O3}

\definecolor{O4}{RGB}{0,255,0}
\def\unifour{O4}

\definecolor{O5}{RGB}{128,0,128}
\def\unifive{O5}

\definecolor{O6}{RGB}{139,69,19}
\def\unisix{O6}

\begin{tikzpicture}[scale=0.8]

\begin{axis}[
    axis lines = left,
    xlabel = Round $t$,
    ylabel = $p_t$,
    legend pos=north east,
    legend cell align = left
]
\addplot [mark=triangle*, \unitwo, line width=1.1pt] table[x=round, y=P1, col sep=comma] {\datagraphpvst};
\addlegendentry{$p_{1} = 0$}

\addplot [mark=diamond*, \unithree, line width=1.1pt] table[x=round, y=P2, col sep=comma] {\datagraphpvst};
\addlegendentry{$p_{1} = 0.2$}

\addplot [mark=square*, \unifour, line width=1.1pt] table[x=round, y=P3, col sep=comma] {\datagraphpvst};
\addlegendentry{$p_{1} = 0.4$}

\addplot [mark=o, \unicolor, line width=1.1pt] table[x=round, y=P4, col sep=comma] {\datagraphpvst};
\addlegendentry{$p_{1} = 0.6$}

\addplot [mark=star, \unifive, line width=1.1pt] table[x=round, y=P5, col sep=comma] {\datagraphpvst};
\addlegendentry{$p_{1} = 0.8$}

\addplot [mark=x, \unisix, line width=1.1pt] table[x=round, y=P6, col sep=comma] {\datagraphpvst};
\addlegendentry{$p_{1} = 1$}

\end{axis}

\end{tikzpicture}
\caption{Demonstrating contraction for the instance in  Example~\ref{example}.}
\label{fig:subfig1}
\end{subfigure}
\hfill
\begin{subfigure}[t]{0.49\textwidth}
\centering

\def\datagraphnocontraction{figures/pvst_nocont.csv}

\definecolor{O2}{RGB}{0,0,255}
\def\unitwo{O2}

\definecolor{O1}{RGB}{255,0,0}
\def\unicolor{O1}

\definecolor{O3}{RGB}{0,0,0}
\def\unithree{O3}

\definecolor{O4}{RGB}{0,255,0}
\def\unifour{O4}

\definecolor{O5}{RGB}{128,0,128}
\def\unifive{O5}

\definecolor{O6}{RGB}{139,69,19}
\def\unisix{O6}

\begin{tikzpicture}[scale=0.8]

\begin{axis}[
    axis lines = left,
    xlabel = Round $t$,
    ylabel = $p_t$,
    legend style={at={(1,0.7)}},
    legend cell align = left
]

\addplot [mark=triangle*, \unitwo, line width=1.1pt] table[x=round, y=P1, col sep=comma] {\datagraphnocontraction};
\addlegendentry{$p_{1} = 0.5$}

\addplot [mark=diamond*, \unithree, line width=1.1pt] table[x=round, y=P2, col sep=comma] {\datagraphnocontraction};
\addlegendentry{$p_{1} = 0.6$}

\addplot [mark=square*, \unifour, line width=1.1pt] table[x=round, y=P3, col sep=comma] {\datagraphnocontraction};
\addlegendentry{$p_{1} = 0.7$}

\addplot [mark=o, \unicolor, line width=1.1pt] table[x=round, y=P4, col sep=comma] {\datagraphnocontraction};
\addlegendentry{$p_{1} = 0.8$}

\addplot [mark=star, \unifive, line width=1.1pt] table[x=round, y=P5, col sep=comma] {\datagraphnocontraction};
\addlegendentry{$p_{1} = 0.9$}

\addplot [mark=x, \unisix, line width=1.1pt] table[x=round, y=P6, col sep=comma] {\datagraphnocontraction};
\addlegendentry{$p_{1} = 1$}

\end{axis}

\end{tikzpicture}
\caption{Demonstrating expansion for the instance in Example~\ref{example:contraction}.}
\label{fig:subfig2}
\end{subfigure}
\caption{The two sub-figures showcase contraction and expansion of long-term proportions with varying initial proportions $p_1$. In Figure~\ref{fig:subfig1}, strong contraction applies. The proportion $p_3$ for $t=3$ is almost invariant to $p_1$. In Figure~\ref{fig:subfig2}, the initial proportion determines whether the long-term proportion converges to 0 or 1. This showcases that even an $\varepsilon$-close estimate can be unfruitful to the regulator.
}
\label{fig:combined}
\end{figure*}
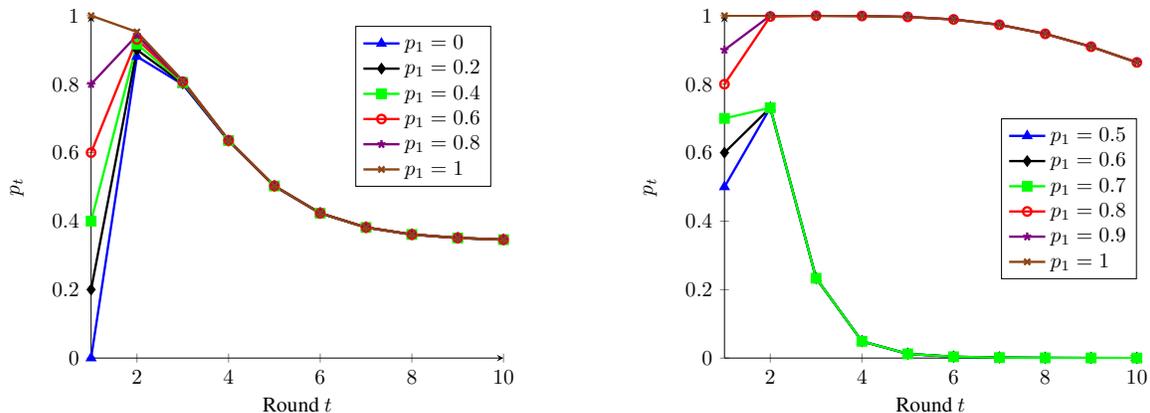

An immediate consequence of Lemma~\ref{prop greater p0} is that
\begin{corollary}\label{cor: bounds with q}
For every $t\in \{0, \dots \Delta -1 \}$, it holds that
\begin{equation}\label{eq:sandwich p}
  q^0_t  \leq p_{\tau+t}(\bm x) \leq q^1_t.  
\end{equation}
\end{corollary}
Notice that the left term of Inequality~\eqref{eq:sandwich p} is obtained by selecting $\alpha = 0$, while the right term is obtained by selecting $\alpha = 1$. Using the bounds in Corollary~\ref{cor: bounds with q}, we can sandwich the utility $u_{\tau+t}(\bm x)$ for every $t\in \{0, \dots \Delta -1 \}$ by 
\begin{align}\label{eq:u sandwithc}
&\unu_{t} = q^0_t \rg(t) + (1-q^1_t)\ro(q^1_t), \nonumber\\
&\ovu_{t} = q^1_t \rg(t) + (1-q^0_t)\ro(q^0_t).     
\end{align}
Indeed, the reader can verify that $\unu_{t} \leq u_{\tau+t}(\bm x) \leq \ovu_{t} $. Equipped with Inequality~\eqref{eq:u sandwithc}, we are ready to give necessary and sufficient conditions.
\begin{theorem} \label{thm: sufficient helpful}
Fix an instance and a scheme $\bm x$, and let $\Delta$ be the maximal number of rounds between two consecutive training rounds. 
\begin{itemize}
    \item Sufficient condition: If $\sum_{t = 0}^{\Delta - 1} \unu_{t} \geq \Delta \ro(0)$ holds, then $\bm x$ is socially beneficial.
    \item Necessary condition: If $\bm x$ is socially beneficial, then $\sum_{t = 0}^{\Delta - 1} \ovu_{t} \geq \Delta \ro(0)$ holds.
\end{itemize}
\end{theorem}
Theorem~\ref{thm: sufficient helpful} provides the regulator with powerful tools. If the regulator is proactive, it can restrict \llm to schemes that satisfy the sufficient condition. By doing so, it ensures that \llm is socially beneficial, at the expense of a revenue decrease for \llm. Alternatively, if the regulator is more passive and only wants to intervene when \llm is guaranteed to be socially harmful, it could act as soon as the necessary condition is violated.

\subsection{Access to Noisy Estimates}\label{subsec:noisy}
While Theorem~\ref{thm: sufficient helpful} provides a practical approach, it does not quantify how far the bounds are from the actual utility. Specifically, the upper bound $\sum_{t = 0}^{\Delta - 1} \ovu_{t}$ and the lower bound $\sum_{t = 0}^{\Delta - 1} \unu_{t}$ could be distant from the actual utility $\sum_{t = 0}^{\Delta - 1}$$ u_{\tau+t}(\bm x)$. In this subsection, we provide a remedy. Throughout this subsection, we limit our attention to instances where $\ro$ is $L$-Lipschitz satisfying $\ro(1) = 0$ and $\beta L < \frac{16}{7}$.

Recall that we have previously assumed that the regulator is not informed about the proportions at all. However, in some cases, the regulator might have access to noisy estimates of the proportions. For instance, the regulator could survey the population and form a confidence interval on the proportion. Next, we shall assume that for every $t \in [T]$, the regulator can construct an estimate $\hat p$ such that $\abs{p_t(\bm x)-\hat p} <\varepsilon$ for $\varepsilon > 0$. Furthermore, we assume $\varepsilon$ is known to the regulator, i.e., it knows the size of the confidence interval.

In such a case, we can complement our monotonicity result (Lemma~\ref{prop greater p0}). The next theorem shows that the Markovian update of proportions forms a contraction mapping.
\begin{theorem} \label{thm: contracting}
Fix any instance and scheme $\bm x$. Let $\varepsilon>0 $ such that $\varepsilon < \frac{16-7\beta L}{14 \beta L e^{\beta L}}$. There exists $\gamma=\gamma(\beta,L,\epsilon)$, $\gamma \in (0,1)$ such that if $\abs{p_{\tau}(\bm x) - q^\alpha_0} < \varepsilon$, then for every $t\in \{0, \dots \Delta -1 \}$ it holds that $\abs{p_{\tau+t}(\bm x) - q^\alpha_t} < \varepsilon \cdot \gamma^t$.
\end{theorem}
Theorem~\ref{thm: contracting} implies that we expect the upper (or lower) bound $q^\alpha_t$ to approach the actual proportion $p_{\tau+t}$ at an exponential rate.
\begin{example}\label{example:contraction}
\normalfont
Recall the instance in Example~\ref{example} and the no-training scheme $\bm x^0$. Figure~\ref{fig:subfig1} demonstrates contraction for $p_t(\bm x)$ assuming different initial proportions $p_1$. After three rounds, the resulting $p_t(\bm x^0)$ is indistinguishable. 

In contrast, consider the instance $\rg(t) = 1.1\cdot 0.8^t$, $\ro(p) = \frac{1}{1+e^{-100 \left( 0.8 - p \right)}}$, $\beta = 10$. Figure~\ref{fig:subfig2} examines how $p_t(\bm x^0)$ changes over time for various selections of $p_1$. This instance violates the condition $\beta L > \frac{16}{7}$, resulting in divergence. Specifically, for $p_1\in [0,0.7]$, the proportion in later rounds convergence to 0. For $p_1 \in [0.8,1]$, the proportion converges to 1. Importantly, two close initial proportions in the range $(0.7,0.8)$ could converge to different proportions in later rounds. 
\end{example}
From here, we use Theorem~\ref{thm: contracting} and the regulator's information $\hat p$ and $\varepsilon$ to provide much tighter necessary and sufficient conditions. For every $t\in \{0,\dots, \Delta-1\}$, instead of using the crude lower bound $q^0_t$ on $p_{\tau+t}(\bm x)$ from Corollary~\ref{cor: bounds with q}, we use the more tighter bound $q^{\hat p-\varepsilon}_t$. Indeed, by the way we define the auxiliary sequence $q$, if $\hat p-\varepsilon \leq p_\tau(\bm x)$ then $q^{\hat p-\varepsilon}_0 \leq p_\tau(\bm x)$; thus, Lemma~\ref{prop greater p0} ensures that $q^{\hat p-\varepsilon}_t \leq p_{\tau+t}(\bm x)$ for every $t\in \{0,\dots, \Delta-1\}$. We form a similar upper bound  $q^{\hat p+\varepsilon}_t \geq p_{\tau+t}(\bm x)$ since $\hat p+\varepsilon \geq p_\tau(\bm x)$. To enhance readability, we use the shorter notation $\unq_t$ for the lower bound, i.e., $\unq_t = q^{\hat p-\varepsilon}_t$, and $\ovq_t = q^{\hat p-\varepsilon}_t$ for the upper bound. Therefore, we know that for every $t\in \{0,\dots, \Delta-1\}$, $\unq_t\leq p_{\tau+t}(\bm x) \leq \ovq_t$. Using these bounds on the proportion, we improve the previously proposed bounds on the instantaneous welfare from Equation~\eqref{eq:u sandwithc},
\begin{align}\label{eq:new u sandwithc}
&\unu^{\varepsilon}_{t} = \unq_t \rg(t) + (1-\ovq_t)\ro(\ovq_t), \nonumber\\
&\ovu^{\varepsilon}_{t} = \ovq_t \rg(t) + (1-\unq_t)\ro(\unq_t).     
\end{align}
As before, we have $\unu^{\varepsilon}_{t} \leq u_{\tau+t}(\bm x) \leq \ovu^{\varepsilon}_{t} $. Using Theorem~\ref{thm: contracting}, we show the following proposition.
\begin{proposition}\label{prop:tight bound on u}
Fix any instance and scheme $\bm x$. Let $\varepsilon > 0$ such that $\varepsilon < \frac{16-7\beta L}{28 \beta L e^{\beta L}}$. There exists $\gamma=\gamma(\beta,L,\epsilon)$, $\gamma \in (0,1)$ such that for every $t\in \{0,\dots, \Delta-1\}$, it holds that 
\begin{align*}
&0\leq u_{\tau+t}(\bm x) - \unu^{\varepsilon}_{t} \leq 2\varepsilon \gamma^t \left( \rg(t) +2L \right), \\
&0 \leq \ovu^{\varepsilon}_{t} - u_{\tau+t}(\bm x) \leq 2\varepsilon \gamma^t \left( \rg(t) +2L \right).
\end{align*}
\end{proposition}
The next theorem combines all the results of this section.
\begin{theorem}\label{thm:u star is not far}
Fix any instance and scheme $\bm x$. Let $\varepsilon > 0$ such that $\varepsilon < \frac{16-7\beta L}{28 \beta L e^{\beta L}}$. There exists $\gamma=\gamma(\beta,L,\epsilon)$, $\gamma \in (0,1)$ such that 
\begin{align}\label{eq: u star bound thm}
\sum_{t = 0}^{\Delta - 1}\ovu^{\varepsilon}_{t}  - \sum_{t = 0}^{\Delta - 1}\unu^{\varepsilon}_{t} \leq 4\varepsilon \sum_{t = 0}^{\Delta - 1} \gamma^t \left( \rg(t) + 2L  \right).
\end{align}
\end{theorem}
The theorem assists the regulator in several ways. First, imagine a proactive regulator wishes to enforce the sufficient condition of Theorem~\ref{thm: sufficient helpful} using the improved bounds $\unu^{\varepsilon}_{t}$, namely to require that $\sum_{t = 0}^{\Delta - 1} \unu^{\varepsilon}_{t} \geq \Delta \ro(0)$. In such a case, the regulator could argue that the margin between $\sum_{t = 0}^{\Delta - 1} \unu^{\varepsilon}_{t}$ and the actual welfare $\sum_{t = 0}^{\Delta - 1} u_{t\tau+t}(\bm x)$ is small; hence, this sufficient condition is not too stringent. Second, imagine a passive regulator that intervenes only if it \emph{knows} that \llm is socially harmful, that is, if $\sum_{t = 0}^{\Delta - 1} \ovu^{\varepsilon}_{t} < \Delta \ro(0)$. The passive regulator is guaranteed that if it does not intervene, the extent to which \llm could be harmful is less than the right-hand-side of Inequality~\eqref{eq: u star bound thm}. 

\section{Discussion and Future Work}\label{sec:discussion}
This paper initiates the research on the dynamics of the competition between generative AI and human-based platforms. After introducing the formal model, we studied \llm's revenue-maximization problem. We have proposed an approximately optimal algorithm, and showed that optimal schemes are not cyclic. Then, we analyze social welfare. We demonstrated a Braess's paradox-like phenomena, where despite that \llm is initially socially beneficial, its impact on \forum's network effects leads to deteriorated welfare. We also showed an infinite Price of Anarchy. Finally, we developed tools that could assist regulators to ensure that GenAI is socially beneficial in the long term without the need to use any proprietary information privately known to \llm.

We see considerable scope for future work. From a technical perspective, some of our results apply to sub-classes of instances. For example, the statements in Subsection~\ref{subsec:noisy} are relevant in cases where $\beta L \leq \frac{16}{7}$. We suspect that this is a by-product of our analysis and the techniques we use to show contraction. Future work could overcome this limitation. More conceptually, our model assumes that \forum is non-strategic. Future work could relax this assumption, modeling \forum as a strategic player, e.g., a data seller, relating to works on information markets~\citep{bergemann2019markets,chen2018optimal,chen2020selling}. In such a case, \forum could strategically decide what to sell, when to sell, and at which price, thereby affecting the welfare dynamics.

\bibliographystyle{abbrvnat}
\bibliography{ms}

{\ifnum\Includeappendix=1{
\onecolumn
\appendix

\section{Transition Functions} \label{sec: transition functions}

In this section, we define the set of \emph{transition functions}, i.e., a set of functions that map the proportion $p_t(\bm x)$ to proportion $p_{t+1}(\bm x)$. Formally, let $(f_t)_t$ be the set of transition functions where $f_t : [0, 1] \rightarrow [0, 1]$ defined as
\[
f_t(p) = \frac{e^{\beta \rg(t)}}{e^{\beta \rg(t)} + e^{\beta \ro(p)}}.
\]

Notice that for any training scheme and any round $t$ it holds that
\begin{align*}
p_{t+1}(\bm x) = \begin{cases}
    f_t(p_t(\bm x)) & \mbox{$x_t = 0$}, \\
    f_0(p_t(\bm x)) & \mbox{$x_t = 1$}.
\end{cases}
\end{align*}

Further, we define the set of \emph{extended transition functions}, namely $(F_t)_t$ where $F_t : [0, 1] \rightarrow [0, 1]$. The extended transition functions map a proportion in round $0$ to a proportion in a future round $t' > t$ using a composition of the transition functions from $t$ to $t'$. Formally $F_t(p) = f_{t-1} \circ \ldots \circ f_1 \circ f_0(p)$.

Throughout the following sections, we show that transition functions posses contraction and monotonic properties.

\section{Proofs Omitted from Section~\ref{llm opt section}}

\begin{proofof}{prop:dynamic}

We first present the algorithm, explain it and then prove our proposition. Throughout this proof, we omit the maintenance cost as it is constant and does not affect the optimization problem.

\paragraph{Part 1: Algorithm presentation}
For convenience, we define $\floor{x}_\varepsilon = \floor{\frac{x}{\varepsilon}} \varepsilon$ the $\varepsilon$-discrete value of $x$.

\begin{algorithm}[t]
\textbf{Input:} $\rvg, c_m, c_{train}, \rg, \ro, \beta, p_1, T, \varepsilon$. \\
\textbf{Output:} $\bm x$
\begin{algorithmic}[1]
\small % add this command to decrease the font size
\caption{Approximate revenue maximizing scheme (ARMS)} \label{alg: ARMS}

\STATE let $M$ a $\frac{1}{\varepsilon} \times (T+1) \times T \times 2$ matrix \label{Line: def m}
\STATE initialize $M(i, t, s, j) = 0$ for every $i \in [\frac{1}{\varepsilon}]$, $t \in [T+1]$, $s\in \{0, 1, ..., T\}$, $j \in \{1, 2\}$. \label{Line: init M}

\FOR{$t = \{T, T-1, \ldots, 1\}$} \label{Line: itr horizon}
    \FOR{$s \in [T-1]$} \label{Line: itr gamma}
        \FOR{$i \in \floor{\frac{1}{\varepsilon}}$} \label{Line: proportion}
            \STATE let $p_{curr} \gets i\varepsilon$ \label{Line: init pcurr}
            \STATE let $p_{next} \gets \floor{\frac{e^{\beta \rg(s)}}{e^{\beta \rg(s)} + e^{\beta \ro(p_{curr})}}}_{\varepsilon}$ \label{Line: init pnext}
            \STATE let $V \gets \left(\rvg p_{curr} + M(p_{next}, t+1, s+1, 1), \rvg p_{curr} - c_{train} + M(p_{next}, t+1, 0, 1)\right)$ \label{Line: init possible rev}
            \STATE $M(p_{curr}, t, s, 1) = \max{V}$ \label{Line: set max rev}
            \STATE $M(p_{curr}, t, s, 2) = \argmax{V}$ \label{Line: set action}
        \ENDFOR
    \ENDFOR
\ENDFOR

\STATE extract $\bm x$ from $M$ \label{line: get scheme}
\RETURN{$\bm x$} \;
\end{algorithmic}
\end{algorithm}

\begin{algorithm}[t]
\textbf{Input:} $M, p_1, T, \varepsilon$. \\
\textbf{Output:} $\bm x$
\begin{algorithmic}[1]
\small % add this command to decrease the font size
\caption{Scheme extraction} \label{alg: extracts}

\STATE Let $\gamma \gets 0$
\FOR{$t = [T]]$} \label{Line: itr horizon}

    \STATE Let $x_t \gets M(p, t, \gamma, 2)$
    \IF{$x_t = 1$}
        \STATE $\gamma \gets 0$
    \ELSE
        \STATE $\gamma \gets \gamma + 1$
    \ENDIF

    \STATE let $p \gets \floor{\frac{e^{\beta \rg(\gamma)}}{e^{\beta \rg(\gamma)} + e^{\beta \ro(p)}}}_{\varepsilon}$ \label{Line: update props}
\ENDFOR

\RETURN{$\bm x$} \;
\end{algorithmic}
\end{algorithm}

Algorithm~\ref{alg: ARMS} approximates the optimal revenue-maximizing training scheme dynamic programming approach \citep{martello1990knapsack}. The algorithm starts at $t = T$ and uses backward induction to solve the problem at $t = 1$. Every element of the matrix $M(i, t, s, 1)$ represents the approximately optimal revenue for the sub-problem with the parameters: $p'_1 = i\varepsilon$, $T' = T-t+1, \gamma'_1 = s$. In each step of the induction, given a proportion $p \in \{0, \varepsilon, 2\varepsilon, 1-\varepsilon \}$, we calculate the next proportion $p_{next}$ and possible revenues by summing the current revenue $\rvg p$ with two possible future revenues. The first revenue represents the case where \llm does not train, while the second possible revenue represents the case where \llm choose to train, namely $M(p_{next}, t+1, 0,1)$.

We further elaborate on the algorithm by explaining each line. Line~\ref{Line: def m} defines the matrix in which we save the calculations of the dynamic programming, while Line~\ref{Line: init M} initialize each element of $M$ with $0$. Line~\ref{Line: itr horizon} formally begins the dynamic programming calculation by iterating over the possible horizons $t \in [T]$, where each $t$ represents a sub-problem with $T' = T - t + 1$ rounds. Next, Line~\ref{Line: itr gamma} iterates over the possible values of $\gamma_t$. The last for loop in Line~\ref{Line: proportion} iterates over the discretization of $[0, 1]$, which represents a finite set of possible proportions. Line~\ref{Line: init pcurr} set the current proportion according to the discretization index in Line~\ref{Line: proportion}. Next, Line~\ref{Line: init pnext} calculates the next $\varepsilon$-discrete proportion, based on the possible number of rounds since the last training round $\gamma_{t} = s-1$ and the proportion $p_curr$. Line~\ref{Line: init possible rev} calculates the two possible revenues. The first revenue $\rvg p_{curr} + M(p_{next}, t+1, s+1, 1)$ represents the case where \llm choose not to train in round $t$, for which the number of round since the last training round increases by 1, i.e $\gamma_{t} = s+1$. The second revenue $\rvg p_{curr} - c_{train} + M(p_{next}, t+1, 0, 1)$ represents the case where \llm choose to train in round $t$, therefore incurring a cost of $c_train$. Furthermore, since \llm train, the number of round since the last training round goes back to $0$, namely $\gamma_{t} = 0$. We finish the backward induction with Lines~\ref{Line: set max rev} and Line~\ref{Line: set action} where Line~\ref{Line: set max rev} updates the maximal attainable revenue and Line~\ref{Line: set action} updates the action ($0$ in case \llm does not train and $1$ it case it does) that leads to that revenue. The final line of our algorithm, Line~\ref{line: get scheme}, extracts the approximately optimal training scheme by recording $M(\floor{p_1}_\varepsilon, 1, 1, 2)$ and following according to the next proportion with the selected action.

\paragraph{Part 2: Proposition proof}

We define a new type of instance to prove our proposition. Let $\tilde{I}(\gamma,T, p) = \tupbracket{\rvg, c_m, c_{train}, \rg, \ro, \beta, p, \gamma, T}$ be an instance of a new problem with $T$ rounds, with the following differences:
\begin{enumerate}
\item \llm is not obliged to train at $t = 1$.
\item \llm train at round $t = 1-\gamma$.
\end{enumerate}

We can think about this problem as the case where \llm train at some point, but is released after $\gamma$ rounds to the users. We define $\tilde{V}(\gamma, T, p)$ the average revenue induced by the optimal training scheme in problem $\tilde{I}(\gamma, T, p)$. Furthermore, let $\tilde{v}_t(\gamma, T, p)$ be the revenue in round $t$ and let $\tilde{p}_t(\gamma, T, p)$ denote the proportion in round $t$ under the optimal training scheme in $\tilde{I}(\gamma, T, p)$. Lastly, we denote $\tilde{\gamma}_t(\gamma, T, p)$ the number of rounds from the last training round relative to round $t$ according to the optimal scheme in instance $I(\gamma, T, p)$. We note the following properties:
\begin{enumerate}
\item $\tilde{V}(\gamma, T, p) = \sum_{t = 1}^T \tilde{v}_t(\gamma, T, p)$.
\item If \llm choose to train in round $t$ then
\[
\tilde{V}(\gamma, T, p) = \sum_{t' = 1}^{t-1} \tilde{v}_{t'}(\gamma, T, p) + \tilde{V}(0, T-t+1, \tilde{p}_t(\gamma, T, p)).
\]
If \llm choose not to train in round $t$ then
\[
\tilde{V}(\gamma, T, p) = \sum_{t' = 1}^{t-1} \tilde{v}_{t'}(\gamma, T, p) + \tilde{V}(\tilde{\gamma}_t(\gamma, T, p), T-t+1, \tilde{p}_t(\gamma, T, p)).
\]
\item Let $\bm x$ be the optimal training scheme, then $\tilde{V}(0, T, p_1) = TV(\bm x)$
\end{enumerate}

The following lemma defines the approximation of each element in $M$.
\begin{lemma} \label{lemma m approx}
For every $t \in [T]$ and every $p \in [p_t(\bm x) - (t-1)\varepsilon, p_t(\bm x) + (t-1)\varepsilon]$ it holds that
\[
\tilde{V}(\gamma, T-t+1, p) - M(\floor{p}_\varepsilon, t, \gamma, 1) < \rvg \varepsilon \sum_{t' = t}^T t'. 
\]
\end{lemma}
Together with the third property, we get
\[
TV(\bm x) - M(p_1(\bm x), 1, 0, 1) < r\varepsilon \sum_{t' = 1}^T t' < \rvg \varepsilon \sum_{t' = 1}^T T = \rvg \varepsilon T^2;
\]
and thus
\[
\frac{1}{T} M(p_1(\bm x), 1, 0, 1) > V(\bm x) - \rvg \varepsilon T.
\]

Let $\bm x'$ be the induced scheme in $M$ according to Algorithm~\ref{alg: extracts}, and let $(q_t)_t$ be the discretized proportions following $\bm x'$ in Algorithm~\ref{alg: extracts}, Line~\ref{Line: update props}. Since $p_1(\bm x') \geq \floor{p_1}_\varepsilon = q_1$, then from Lemma~\ref{prop greater p0} it holds that $p_t(\bm x') \geq q_t$ and $V(\bm x') \geq \frac{1}{T} M(p_1(\bm x), 1, 0, 1)$. Finally, to conclude our results so far
\[
V(\bm x') \geq \frac{1}{T} M(p_1(\bm x), 1, 0, 1) > V(\bm x) - \rvg \varepsilon T.
\]
\end{proofof}

\begin{proofof}{lemma m approx}
We prove by backward induction over $t$, starting with $t = T$ and then prove for $t < T$. We start with the base case $t = T$.  The optimal revenue can be expressed as
\[
\tilde{V}(\gamma, 1, p) = \tilde{v}(\gamma, 1, p) = \max_{x_1} p \rvg - x_1 c_{train} = p \rvg.
\]

According to our algorithm, at $t = T+1$, matrix $M$ satisfies $M(i, T+1, s, 1) = 0$ for every $i \in \ceil{\frac{1}{\varepsilon}}$ and $s \in [T]$. Therefore, for every $i \in \ceil{\frac{1}{\varepsilon}}$, the approximated revenue is
\begin{align*}
M(\floor{p}_\varepsilon, T, \gamma, 1) &= \floor{p}_\varepsilon \rvg + \max\{-c_{train} + M(i, T+1, \gamma + 1, 1), M(i, T+1, 0, 1)\} \\
&= \floor{p}_\varepsilon \rvg + \max\{-c_{train} + 0, 0\} = \floor{p}_\varepsilon \rvg.
\end{align*}

Therefore, the difference in the revenue is
\begin{align*}
\tilde{V}(\gamma, 1, p) - M(\floor{p}_\varepsilon, T, \gamma_{T}, 1) &= \tilde{v}(\gamma, 1, p) - M(\floor{p}_\varepsilon, T, \gamma_{T}, 1) \\
&= p \rvg - \floor{p}_\varepsilon \rvg < \rvg T\varepsilon.
\end{align*}
We finished with the base case. Now, assume the lemma is true for $t+1$, and we prove it for $t$.
Let $s = \gamma$. Then, according to the second property, if \llm choose to train at $t = 1$
\[
\tilde{V}(\gamma, T-t+1, p) = \tilde{v}_1(\gamma, T-t+1, p) + \tilde{V}(0, T-t, \tilde{p}_2(\gamma, T-t+1, p)),
\]
and if \llm choose not to train at $t = 1$ then
\[
\tilde{V}(\gamma, T-t+1, p) = \tilde{v}_1(\gamma, T-t+1, p) + \tilde{V}(\tilde{\gamma}_2(\gamma, T-t+1, p), T-t, \tilde{p}_2(\gamma, T-t+1, p)).
\]
We now calculate the next proportion according to Line~\ref{Line: init pnext}. Let $p' = \frac{e^{\beta \rg(s)}}{e^{\beta \rg(s)} + e^{\beta \ro(\floor{p}_{\varepsilon})}}$ and notice that $p_{next} = \floor{p'}_{\varepsilon}$. We use Theorem~\ref{thm: contracting} to estimate the distance between $\tilde{p}_2(\gamma, T-t+1, p)$ and $p'$. First, notice that $\abs{p - \floor{p}_\varepsilon} < t\frac{16-7\beta L}{14 \beta L e^{\beta L} T} = t \varepsilon < \frac{16-7\beta L}{14 \beta L e^{\beta L}}$

\begin{align*}
\left| \tilde{p}_2(\gamma, T-t+1, p) - p' \right| < t\varepsilon.
\end{align*}

Therefore, $p' \in [\tilde{p}_2(\gamma, T-t+1, p) - t\varepsilon, \tilde{p}_2(\gamma, T-t+1, p) + t\varepsilon]$. Hence, by the assumption of the induction, it holds that
\begin{enumerate}
    \item $\tilde{V}(0, T-t, \tilde{p}_2(\gamma, T-t+1, p)) - M(\floor{p'}_\varepsilon, t+1, 0, 1) < \rvg \varepsilon \sum_{t' = t+1}^T t'$.
    \item $\tilde{V}(\tilde{\gamma}_2(\gamma, T-t+1, p), T-t, \tilde{p}_2(\gamma, T-t+1, p)) - M(\floor{p'}_\varepsilon, t+1, \tilde{\gamma}_2(\gamma, T-t+1, p), 1) < \rvg \varepsilon \sum_{t' = t+1}^T t' $.
\end{enumerate}
Thus, we can lower bound $M(\floor{p}_\varepsilon, t, s, 1)$ by
\begin{align*}
M(\floor{p}_\varepsilon, t, s, 1) > &\rvg \floor{p}_\varepsilon - \rvg \varepsilon \sum_{t' = t+1}^T t' \\
&+ \max \{ \tilde{V}(0, T-t, \tilde{p}_2(\gamma, T-t+1, p)), \tilde{V}(\tilde{\gamma}_2(\gamma, T-t+1, p), T-t+2, \tilde{p}_2(\gamma, T-t+1, p)) \}.
\end{align*}
Finally, the difference in the future revenue is
\begin{align*}
V(\gamma, T-t+1, p) - M(\floor{p}_\varepsilon, t, \gamma, 1) &< \rvg p - \rvg \floor{p}_\varepsilon + \rvg \varepsilon \sum_{t' = t+1}^T t' \\
&\leq \rvg \left( t \varepsilon + \varepsilon \sum_{t' = t+1}^T t' \right) \\
&= \rvg \varepsilon \sum_{t' = t}^T t'
\end{align*}

\end{proofof}

\begin{proofof}{prop: proportions converge}
Recall the definition of transition functions from Section~\ref{sec: transition functions}. Therefore, notice that for every $i \in [k]$ it holds that

\begin{align*}
p_{l\cdot k +i}(\bm x_T^k) = \begin{cases}
    f_{k-1}(p_{l\cdot k }(\bm x_T^k)) & \mbox{$i = 1$}, \\
    f_{i-2}(p_{l\cdot k +i-1}(\bm x_T^k)) & \mbox{$i > 1$}.
\end{cases}
\end{align*}

Therefore, we can define an extended transition function that maps proportion $p_{l\cdot k +i}(\bm x_T^k)$ to $p_{(l+1)\cdot k +i}(\bm x_T^k)$ using a composition of $\{ f_i \}_{i=0}^{k-1}$.

\begin{align*}
    p_{(l+1)\cdot k +i}(\bm x_T^k) = \begin{cases}
       f_{k-1} \circ \ldots \circ f_1 \circ f_0(p_{l\cdot k + 1}(\bm x_T^k)) = F_k(p_{l\cdot k + 1}(\bm x_T^k)) & \mbox{$i = 1$}, \\
       f_{i-2} \circ \ldots \circ f_0 \circ f_{k-1} \circ \ldots \circ f_{i-1}(p_{l\cdot k + i}(\bm x_T^k)) & \mbox{$i > 1$}
    \end{cases}
\end{align*}

\begin{proposition} \label{composition monotone}
Let $F(p)$ be a composition of functions in $\{f_t\}_t$. Then, $F(p)$ is differentiable and satisfies $\frac{d}{dp}F(p) \geq 0$ for every $p \in [0, 1]$.
\end{proposition}

Proposition~\ref{composition monotone} suggests that $F(p)$ is bounded and monotone from $[0, 1]$ to $[0, 1]$, therefore by monotonic sequence theorem, the sequence $(p_{l\cdot k +i}(\bm x_T^k))_l$ converges
\end{proofof}

\begin{proofof}{composition monotone}
We first prove that the derivative of $f_t(p)$ is non-negative, that is
For every $t \in [T-1]$, $f_t(p)$ satisfies $\frac{d}{dp}f_t(p) \geq 0$. By definition
\[
\frac{d}{dp}f_t(p) = -\beta f_t(p)(1-f_t(p)) \frac{d}{dp}\ro(p) \geq 0.
\]
The inquality holds since $f_t(p) \geq 0$, $\beta \geq 0$ and from assumption $\frac{d}{dp}\ro(p) < 0$.

Next, let superscript $^i$ denote the order of functions in the composition, that is $F(p) = f^{k} \circ \ldots \circ f^{2} \circ f^{1}(p)$.
Therefore, the derivative of $F(p)$ using the chain rule is:
\begin{align*}
\frac{d}{dp}F = \frac{df^k}{df^{k-1}} \frac{df^{k-1}}{df^{k-2}} \ldots \frac{df^{2}}{df^{1}} \frac{df^{1}}{dp} \geq 0.
\end{align*}
\end{proofof}

\begin{proofof}{thm not cyclic}
Consider the instance: $\rvg = 1$, $c_m = 0.6$, $c_{train} = 0.504$, $\rg(t) = 3(0.5)^t$, $\ro(p) = 1-p$, $\beta = 1$, $p_0 = 1$. To prove this theorem, we need to present a (non-cyclic) training scheme that outperforms all cyclic training schemes. The proof consists of 4 steps. In the first step, we extend our definition of cyclic schemes. we show the asymptotic revenue of \llm for any cyclic training scheme and show that the difference from using training schemes with with a prefix and suffix is $\Theta(1/T)$. Then we show that for our instance, the optimal training has to train at most every $K \in \mathbb{N}$ rounds, and therefore, we can limit ourselves to cyclic training schemes with periods in $[K]$. In the third step, we show how to evaluate the asymptotic revenue for every cyclic training scheme. Lastly, we provide a training scheme, which is not cyclic and not necessarily optimal, that achieves a higher asymptotic revenue by at least $\Theta(1)$. Building upon the asymptotic revenue analysis, $T^\star$ is a round for which it is guaranteed that for any horizon $T > T^\star$, the non-cyclic training scheme induces a higher revenue than any other cyclic training scheme.

\paragraph{Step 1: Asymptotic revenue of cyclic training schemes}
We extend our proof to a wider class of training schemes, where the suffix and prefix of the scheme can vary. formally, we define \emph{broadly cyclic} schemes as follows:
\begin{definition}
Training scheme $\bm x$ is \emph{$k$-broadly cyclic} if there exists $t_0, t_1$ such that for every $t \in [t_0, T-t_1]$ it holds that
\begin{align*}
x_t = \begin{cases}
    1 & \mbox{$t - t_0 \pmod k = 0$}, \\
    0 & \mbox{otherwise}.
\end{cases}
\end{align*}
\end{definition}

We show that for every training scheme, the revenue converges when taking the limit of $T \rightarrow \infty$. This limit reflects the average revenue in each period cycle, independent of $t_0$ and $t_1$. It, in turn, allows us to find the optimal cyclic training scheme by comparing the average revenue of a cycle when $T \rightarrow \infty$. Fix $k$ and a $k$-broadly cyclic training scheme.

Recall the definition of transition functions from Section~\ref{sec: transition functions} and notice the following observation:

\begin{observation}\label{obs: transition from pt0}
For every $j \in [k]$ and $i \in \mathbb{N}$ such that $t_0 + ki < T-t_1$ it holds that $p_{t_0 + ki + j}(\bm x^k) = F_j(p_{t_0 + ki}(\bm x^k))$.
\end{observation}

A crucial element in our analysis is the proportion of \llm users at the beginning of the cycle. For that, let $a^k_i = p_{t_0 + k(i-1)}(\bm x^k)$ denote the proportion at the beginning of the $i$'th cycle, for every $i \in \mathbb N$. Observation~\ref{obs: transition from pt0} suggests that $(a_i)_i$ can be defined recursively using $F_k$.

\begin{observation}\label{obs: a is recursive}
We can define $a^k_i$ recursively by $a^k_{i+1} = F_k(a^k_i)$ with the initial value being $a^k_1 = p_{t_0}(\bm x^k)$.
\end{observation}

Notice that the revenue between $t_0$ to $t_1$ can be written as a sum over $\floor{\frac{T - t_0 - t_1}{k}}$ periodic cycles and therefore the revenue of scheme $\bm x^k$ can be written as

\begin{align}\label{eq:redefine rev}
V(\bm x^k) &= \frac{1}{T}\sum_{t = 1}^T v_t(\bm x^k) \nonumber\\
&= \frac{1}{T}\left( \sum_{t = 1}^{t_0} v_t(\bm x^k) + \sum_{t = t_0+1}^{T-t_1} v_t(\bm x^k) + \sum_{t = T-t_1 + 1}^T v_t(\bm x^k) \right) \nonumber \\
&= o\left(\frac{1}{T}\right) + \frac{1}{T} \sum_{t = t_0+1}^{T-t_1} v_t(\bm x^k) \nonumber \\
&= o\left(\frac{1}{T}\right) + \frac{1}{T} \sum_{i = 1}^{\floor{\frac{T - t_0 - t_1}{k}}}\left(\sum_{t = 1}^k p_{t_0 + (i-1)k + t}(\bm x^k) - kc_m - c_{train} \right) \nonumber \\
&=o\left(\frac{1}{T} \right)+\frac{1}{T} \sum_{i = 1}^{\floor{\frac{T - t_0 - t_1}{k}}}\left(\sum_{t = 1}^{k} F_t(a^k_i) - k c_m - c_{train}\right).
\end{align}
While Equation~\eqref{eq:redefine rev} seems complex, it can be significantly simplified using an asymptotic analysis of the sequence $(a^k_i)_i$.
\begin{proposition} \label{composition converge}
There exists $a^\star$ such that $\lim_{t \rightarrow \infty} a_i = a^\star$.
\end{proposition}

From Proposition~\ref{composition converge} we get that there exists $a^{\star,k} \in [0, 1]$ such that $\lim_{t \rightarrow \infty} a_i^k = a^{\star, k}$, Thus
\[
\lim_{T \rightarrow \infty} V(\bm x^k) = \frac{\sum_{t = 1}^{k} F_t(a^{\star, k})  - k c_m - c_{train}}{k} = \frac{\sum_{t = 1}^{k} F_t(a^{\star, k}) - c_{train}}{k} - c_m.
\]

Thus, there exists $T_0$ such that if $T > T_0$ then the optimal cyclic training scheme is the training scheme that satisfies $k = \argmax_k\frac{\sum_{t = 1}^{k} F_t(a^{\star, k}) - c_{train}}{k}$. 
Furthermore, notice that for any two training schemes with the same cycle but different prefix and suffix, the difference is $\Theta(\frac{1}{T})$. Our goal is to find a non-cyclic training scheme $\bm x'$ that achieves $V(\bm x') - \max_k V(\bm x^k) = \Theta(1)$, and therefore from here on, we focus on cyclic training schemes with $t_0 = t_1 = 0$.

\paragraph{Step 2: Bound on the number of cyclic training schemes} Let $\bm x^\star$ be the optimal training scheme and $\tau_i^\star$ be the $i$'th training round of $\bm x^\star$. In this step, we show that there exists $T_0 \in \mathbb{N}$ such that for every $i \in \mathbb{N}$ it holds that $\tau_{i+1}^\star - \tau_i^\star \leq T_0$. Therefore, if the optimal training scheme is cyclic, then the cycle length, representing the number of rounds between two consecutive training rounds, is bounded by $T_0$. Thus, finding the optimal cyclic training scheme is reduced to calculating $T_0$ limits as shown in the step 1.

Let $\bm x^0$ be the training scheme that does not train, defined by 
\begin{align*}
x_t^0 = \begin{cases}
    1 & \mbox{$t = 0$}, \\
    0 & \mbox{Otherwise}.
\end{cases}
\end{align*}

The scheme $\bm x^0$ is an auxiliary scheme which we use to show that if \llm does not train then the revenue becomes negative after certain number of steps. 

\begin{lemma} \label{no train negative revenue}
If there exists $t' \in \mathbb{N}$ such that $p_{t'}(\bm x^0) \geq p_{t'+1}(\bm x^0)$ and the probability $p'$ that satisfies $p' = \frac{e^{\beta \inf_t \rg(t)}}{e^{\beta \inf_t \rg(t)} + e^{\beta \ro(p')}}$ also satisfies $p' < \frac{c_m}{r}$ then there exists $T_0$ such that for every $T > T_0$ it holds that $V(\bm x^0) < 0$.
\end{lemma}

Observe that $p_3(\bm x^0) = 0.81 < 0.95 = p_2(\bm x^0)$. Furthermore, $\inf_t \rg(t) = 0$ and the proportion $p$ satisfying $p = \frac{e^{\beta \inf_t \rg(t)}}{e^{\beta \inf_t \rg(t)} + e^{\beta \ro(0)}}$ is $p = 0.34 < 0.6 = \frac{c_m}{r}$. Therefore, by Lemma~\ref{no train negative revenue}, there exists $T_0$ such that for every $T > T_0$ it holds that $V(\bm x^0) < 0$ and observe that $T_0 = 8$.

Our goal now is to take an extreme case of our instance, that is an equivalent instance with $p_1 = 0$, where there exists another training scheme with training rounds and which results in a positive revenue in $t \geq T_0$ rounds. Combining this training scheme with the monotonicity of the proportions from Lemma~\ref{prop greater p0} leads to the result where the training scheme also induce positive revenue in our original instance. Thus, concluding that there must be a training round in $[t]$.

Denote our instance by $I$ and we define a new instance $I_0 = \{r, c_m, c_{train}, \rg, \ro, \beta, 0 \}$ which is the same as instance $I$, but with an initial probability $p_0 = 0$. Instance $I_0$ is used together with the following theorem to bound the number of rounds between two consecutive training rounds of the optimal training scheme in instance $I$.

\begin{theorem} \label{limited training gap}
Let $(\tau^*_i)$ be the ordered training rounds of the optimal training scheme in instance $I$. If there exists a training scheme $\bm x'$ on instance $I_0$ and $T \geq T_0$ such that $V(\bm x') \geq 0$ then for every $i$ such that $\tau_i^\star \leq T-T_0$ on instance $I$ it holds that $\tau_{i+1}^\star - \tau_{i}^\star \leq T_0$.
\end{theorem}

We now calculate the revenue of scheme $\bm x^3$ on instance $I_0$. Observe that $\sum_{t = 1}^8 v_t(\bm x^3) > 0.397 > 0$. Thus, by theorem~\ref{limited training gap}, the number of rounds between every two consecutive training rounds of $\bm x^\star$ is at most $8$, and therefore we can focus on the training schemes $\bm x^k$ for $k \in \{1,\dots,8\}$.

\paragraph{Step 3: Asymptotic revenue evaluation}
We show a technique to evaluate the asymptotic revenue for any cyclic training scheme. We show our technique for $\bm x^3$. 

Our first goal is to evaluate $a^{\star, k}$. Recall from Equation~\eqref{eq:redefine rev} that for every $i$, the $i$'th cycle include the portions $F_t(a^k_{i})$ for $t \in [k]$. We start with the difference between the portion before the $5'th$ cycle and the $4'th$ cycle, namely $a_5^3$ and $a_4^3$. Afterwards, we show that $a^{\star, 3}$ is a function of $a_5^3 - a_4^3$.

\begin{align} \label{eq: a deff}
&\left| a_5^3 - a_4^3 \right| = \left| p_{13}(\bm x^3) - p_{10}(\bm x^3) \right| < 3.6\cdot 10^{-9}.
\end{align}

Our next goal is to establish the convergence rate of $(a_i)_i$. In fact, looking at the subsequence $(a_i)_{i = 4}^{\infty}$, we show that $F_k(p)$ satisfies a local contraction property, which establishes the convergence rate of $(a_i)_i$, and together with the result of Equation~\ref{eq: a deff}, defines an upper bound for $\left| a^{\star, k} - a^{k}_4 \right|$. Recall that $F_k$ is a composition of the functions $\{f_t \}_{t = 0}^{k-1}$. Therefore, our next two lemmas show that $f_t$ exhibits the contracting property.

\begin{lemma} \label{close probabilities contracting}
Let $p^1, p^2 \in [0,1]$. If $\left| p^1 - p^2 \right| < 0.002$ then for every $t \in [k]$ it holds that
\[
\left| F_t(p^1) - F_t(p^2) \right| < 0.44^t \left| p^1 - p^2 \right|.
\]
\end{lemma}
By applying Lemma~\ref{close probabilities contracting} over $(a_i)$ for every $i > 4$ we get that $(a_i)_{i = 4}^{\infty}$ contracts with a contraction factor of $\gamma^k$.
\begin{corollary}
For every $j > 4$ it holds that
\[
\abs{a_{j+2} - a_{j+1}} < 0.44^k \abs{a_{j+1} - a_{j}}.
\]
\end{corollary}
Therefore, our next proposition uses this contraction property to establish a bound on the limit of the sequence $a^{\star, k}$.

\begin{proposition} \label{opt near vicinity}
Let $p \in [0, 1]$ and let $f : [0,1] \rightarrow [0,1]$. If there exists $p^\star \in [0, 1]$ such that $p^\star = f(p^\star)$ and $f$ satisfies $\left| f^{(n+1)}(p) - f^{(n)}(p)\right| \leq \gamma \left| f^{(n)}(p) - f^{(n-1)}(p) \right|$ for any $n \in \mathbb{N}$ and some $\gamma < 1$ then $\left| p - p^\star \right| < \frac{\left| f(p) - p \right|}{1-\gamma}$.
\end{proposition}

By Proposition~\ref{opt near vicinity} we get
\[
\varepsilon = \left| a^{\star, k} - a^k_4 \right| < \frac{\left| a^k_5 - a^k_4 \right|}{1-0.44^k} = \frac{\left| a^3_5 - a^3_4 \right|}{1-0.44^3} < 3.93 \cdot 10^{-9}.
\]
Since $\left| a^{\star, k} - a^k_4 \right| < 0.002$, we can use it together with Lemma~\ref{close probabilities contracting} to bound the expression $\left| F_t(a^{\star,k}) - F_t(a^{k}_4) \right|$ for every $t \in [k]$, thus formally get
\[
\left| F_t(a^{\star,k}) - F_t(a^{k}_4) \right| < 0.44^t \left| a^{\star,k} - a^{k}_4 \right| < \varepsilon
\]

Finally, we can evaluate how close $\sum_{i = 1}^k F_t(a^k_4)$ is to $\sum_{i = 1}^k F_t(a^{\star, k})$.
\begin{align*}
\left| \sum_{i = 1}^k F_t(a^k_4) - F_t(a^{\star, k}) \right| \leq \sum_{i = 1}^k \left| F_t(a^k_4) - F_t(a^{\star, k}) \right| < k\varepsilon < 1.19 \cdot 10^{-8},
\end{align*}
which means that
\[
\sum_{i = 1}^k F_t(a^k_4) -1.19 \cdot 10^{-8} < \sum_{i = 1}^k F_t(a^{\star, k}) < \sum_{i = 1}^k F_t(a^k_4) + 1.19 \cdot 10^{-8}
\]
Observe that $\sum_{i = 1}^k F_t(a^k_4) = \sum_{t = 10}^{12}p_t(\bm x^k) = 2.37637$ and therefore $\sum_{i = 1}^k F_t(a^{\star, k}) \in (2.3763, 2.3764)$.

We now repeat the same calculation for every $k \in [T_0]$ and show the results in the following table:

\begin{center}
    \begin{tabular}{c|c|c}
    \hline
    Scheme & $\sum_{i = 1}^k F_t(a^{\star, k})$ & $\lim_{T \rightarrow \infty} V(\bm x^k) + c_m$ \\ [0.5ex]
    \hline\hline
        $\bm x^1$ & $(0.9502, 0.9503)$ & $(0.4462, 0.4463)$ \\
        $\bm x^2$ & $(1.752, 1.7522)$ & $(0.624, 0.6241)$ \\
        $\bm x^3$ & $(2.3763, 2.3764)$ & $(0.6241, 0.62413)$ \\
        $\bm x^4$ & $(2.8685, 2.8686)$ & $(0.591, 0.5912)$ \\
        $\bm x^5$ & $(3.2847, 3.2848)$ & $(0.5561, 0.5562)$ \\
        $\bm x^6$ & $(3.6624, 3.6625)$ & $(0.5264, 0.5265)$ \\ 
        $\bm x^7$ & $(4.0213, 4.0214)$ & $(0.5024, 0.5025)$ \\ 
        $\bm x^8$ & $(4.3711, 4.3712)$ & $(0.4833, 0.4834)$ \\ [1ex]
    \hline
    \end{tabular}
\end{center}

Looking at the intervals of $\lim_{T \rightarrow \infty} V(\bm x) + c_m$, we see that the lower bound of $\lim_{T \rightarrow \infty} V(\bm x^3)$ is greater or equal to the upper bound of any other $k \in [T_0]$. Therefore $3 = \argmax_{k} \lim_{T \rightarrow \infty} V(\bm x^k)$.

\paragraph{Step 4: Non-cyclic training scheme} Denote $\bm x^{\alpha_1, \alpha_2}$ the training scheme that alternates its training every $\alpha_1$ and $\alpha_2$ rounds, formally we define 
\begin{align*}
\mathcal{T}^{\alpha_1, \alpha_2} &= \{1 + (N+1)\alpha_1 + N\alpha_2 \mid N \in \mathbb{N}_0\} \cup \{ 1 + N(\alpha_1 + \alpha_2) \mid N \in \mathbb{N}_0\} \\
&= \{1, 1 + \alpha_1, 1 + \alpha_1 + \alpha_2, \ldots  \};
\end{align*}
therefore, the scheme $\bm x^{\alpha_1, \alpha_2}$ is defined by
\begin{align*}
    x_t^{\alpha_1,\alpha_2} = \begin{cases}
        1 & \mbox{$t \in \mathcal{T}^{\alpha_1,\alpha_2}$} \\
        0 & \mbox{otherwise}
    \end{cases}.
\end{align*}

The techniques we employ are similar to step 1 and 3 for scheme $\bm x^{\alpha_1, \alpha_2}$. In this case, we define a period of length $\alpha_1 + \alpha_2$ and look at the proportions every $\alpha_1 + \alpha_2$ rounds. Following step 1, we define the transition function $F_{\alpha_1, \alpha_2}(p) = F_{\alpha_2} \circ F_{\alpha_1}(p)$ and the following sequence $(a^{\alpha_1, \alpha_2})_{i=1}^{\infty}$ such that $a^{\alpha_1, \alpha_2}_{i} = p_{(i-1)}(\bm x^{\alpha_1,\alpha_2})$. Our next observation suggests that $(a^{\alpha_1,\alpha_2}_i)_i$ can be expressed recursively using $F_{\alpha_1,\alpha_2}$.

\begin{observation} \label{obs: general a is recursive}
We can define $a^{\alpha_1,\alpha_2}_i$ recursively by $a^{\alpha_1,\alpha_2}_{i+1} = F_{\alpha_1,\alpha_2}(a^{2,3}_i)$ with the initial value being $a^{\alpha_1,\alpha_2}_1 = p_0$.
\end{observation}

From the same argument as in Step 1, there exists $a^{\star, \alpha_1, \alpha_2}$ such that $\lim_{i \rightarrow \infty} a^{\alpha_1, \alpha_2}_i = a^{\star, \alpha_1, \alpha_2}_i$. Therefore, we can write the asymptotic revenue of \llm under the training scheme $\bm x^{\alpha_1, \alpha_2}$ by
\[
\lim_{t \rightarrow \infty} V(\bm x^{\alpha_1, \alpha_2}) = \frac{\left( \sum_{t = 1}^{\alpha_1} F_t(a^{\star,\alpha_1, \alpha_2}) + \sum_{t = 1}^{\alpha_2} F_t(F_{\alpha_1}(a^{\star, \alpha_1, \alpha_2})) \right) \rvg - c_{train}}{\alpha_1 + \alpha_2} - c_m.
\]
We now focus on a specific training scheme where $\alpha_1 = 2$ and $\alpha_2 = 3$, namely $\bm x^{2,3}$. Notice that Lemma~\ref{close probabilities contracting} and Proposition~\ref{opt near vicinity}, which were used in Step 3, are also applicable for training scheme $\bm x^{2, 3}$, and therefore we employ the same technique as in step 3 to evaluate the expression $\sum_{t = 1}^{\alpha_1} F_t(a^{\star,\alpha_1, \alpha_2}) + \sum_{t = 1}^{\alpha_2} F_t(F_{\alpha_1}(a^{\star, \alpha_1, \alpha_2}))$.

Observe that $\sum_{t = 1}^{\alpha_1} F_t(a^{\star,\alpha_1, \alpha_2}) + \sum_{t = 1}^{\alpha_2} F_t(F_{\alpha_1}(a^{\star, \alpha_1, \alpha_2})) \in (4.1287, 4.1288)$ and $\lim_{t \rightarrow \infty}V(\bm x^{2,3}) + c_m \in (0.62414, 0.62416)$. That is, the lower bound of the asymptotic revenue induced by $\bm x^{2,3}$ is strictly higher than the upper bound of $\bm x^3$, specifically,
\[
\lim_{t \rightarrow \infty} V(\bm x^{2,3}) - \max_{k} \lim_{t \rightarrow \infty} V(\bm x^k) = \lim_{t \rightarrow \infty} V(\bm x^{2,3}) - V(\bm x^3) > 10^{-5};
\]
hence, there exists $T_0$ such that for every $T > T_0$, the scheme $\bm x^{2,3}$ results in higher revenue than the optimal cyclic training scheme $\bm x^3$.
\end{proofof}

\begin{proofof}{obs: transition from pt0}
We show it by induction over $j$. Starting with the base case of $j = 1$. Recall that $\bm x^k$ satisfy $x_t = 1$ for every $t = t_0 + ki$. Therefore, by definition,
\[
p_{t_0 + ki + 1}(\bm x^k) = \frac{e^{\beta \rg(0)}}{e^{\beta \rg(0)} + e^{\beta \ro(p_{t_0 + ki}(\bm x^k))}} = f_0(p_{t_0 + ki}(\bm x^k)) = F_1(p_{t_0 + ki}(\bm x^k)).
\]
Assume the observation holds for every $j < k$ and we show for $i = k$. By definition
\begin{align*}
p_{t_0 + ki + j}(\bm x^k) &= \frac{e^{\beta \rg(j-1)}}{e^{\beta \rg(j-1)} + e^{\beta \ro(p_{t_0 + ki + j-1}(\bm x^k))}} \\
&= f_{j-1}(p_{t_0 + ki + j-1}(\bm x^k)) \\
&= f_{j-1}(F_{j-1}(p_{t_0 + ki}(\bm x^k))) \\
&= f_{j-1} \circ f_{j-2} \dots f_{0}(p_{t_0 + ki}(\bm x^k)) = F_j(p_{t_0 + ki}(\bm x^k)).
\end{align*}
This completes the proof.
\end{proofof}

\begin{proofof}{obs: a is recursive}
We show it by induction over $i$. Starting with the base case of $i = 1$, then by definition $a_1 = p_{t_0 + k(i-1)}(\bm x^k) = p_{t_0}(\bm x^k)$. Assume this is true for any $i$ and we show for $i + 1$. By definition
\begin{align*}
a_{i+1} &= p_{t_0 + ki}(\bm x^k) = p_{t_0 + k(i-1) + k}(\bm x^k) = F_k(p_{t_0 + k(i-1)}(\bm x^k)) = F_k(a_i).
\end{align*}
\end{proofof}

\begin{proofof}{composition converge}
This proposition is a special case of Proposition~\ref{prop: proportions converge}, hence we omit the details.
\end{proofof}

\begin{proofof}{no train negative revenue}
We use the following lemma to analyze the proportions $(p_t)_{t'}^T$.
\begin{lemma} \label{lemma: pt decreasing}
If $p_{t'}(\bm x^0) \geq p_{t' + 1}(\bm x^0)$ then $(p_t(\bm x))_{t = t' + 1}^T$ is a strictly decreasing sequence.
\end{lemma}

Lemma~\ref{lemma: pt decreasing} suggests that there exists $t'' > t'$ such that for every $t > t''$ it holds that
\[
p_t(\bm x^0) < \frac{c_m}{r}.
\]

Let $\varepsilon = p_{t''}(\bm x^0) \rvg - c_m < 0$, and notice that
\[
T V(\bm x^0) = \sum_{t = 1}^{t'' - 1} v_t(\bm x^0) + \sum_{t = t''}^{T} v_t(\bm x^0) < \sum_{t = 1}^{t'' - 1} v_t(\bm x^0) + (T-t'')\varepsilon'
\]
thus, for every $T > \left| \frac{-\sum_{t = 1}^{t'' - 1} v_t(\bm x^0) + t''\varepsilon}{\varepsilon} \right|$ it holds that$V(\bm x^0) < 0$.
\end{proofof}

\begin{proofof}{lemma: pt decreasing}
We prove by induction that if $p_{t - 1}(\bm x) \geq p_{t}(\bm x)$ then $p_{t}(\bm x) > p_{t+1}(\bm x)$ for every $t \geq t' + 1$. We prove the induction over $t$. We start with the base case $t = t' + 1$

From definition of $p_t(\bm x)$ we get $\rg(\gamma_{t-1}) - \ro(p_{t-1}(\bm x)) = \ln\left(\frac{p_t(\bm x)}{1-p_t(\bm x)}\right)$. Notice that $\ln\left(\frac{p}{1-p}\right)$ is a strictly increasing function in $p$, therefore $p_t(\bm x) > p_{t+1}(\bm x)$ if and only if $\ln \left(\frac{p_t(\bm x)}{1-p_t(\bm x)} \right) > \ln \left(\frac{p_{t+1}(\bm x)}{1-p_{t+1}(\bm x)} \right)$.
\begin{align*}
\ln \left(\frac{p_{t+1}(\bm x)}{1-p_{t+1}(\bm x)} \right) - \ln \left(\frac{p_t(\bm x)}{1-p_t(\bm x)} \right) &= 
\rg(\gamma_t) - \ro(p_t(\bm x)) - (\rg(\gamma_{t-1}) - \ro(p_{t-1}(\bm x))) \\
&= \rg(\gamma_t) - \rg(\gamma_{t-1}) + \ro(p_{t-1}(\bm x)) - \ro(p_t(\bm x)).
\end{align*}

Since $\rg$ is a decreasing function we get that $\rg_t - \rg_{t-1} < 0$. From the condition that $p_{t-1}(\bm x) \geq p_t(\bm x) $ we get that $\ro(p_t(\bm x)) \geq \ro(p_{t-1}(\bm x))$, therefore $\ln(\frac{p_{t+1}(\bm x)}{1-p_{t+1}(\bm x)}) - \ln(\frac{p_t(\bm x)}{1-p_t(\bm x)}) < 0$ which ultimately gives us $p_{t+1}(\bm x) < p_t(\bm x)$.

We finished with the base case. The proof of the induction step is the same as the proof of the base case above, and therefore it completes our proof.
\end{proofof}

\begin{proofof}{limited training gap}
Let $I_p = (\rg, \ro, \beta, p, \rvg, c_m, c_{train})$ be the instance with $p_0 = p$. Then from Lemma~\ref{no train negative revenue} for each instance $I_p$ there exists $T(p)$ such that for every $T > T(p)$ it holds that $V(\bm x^0) < 0$. Let $p^p_t(\bm x)$ be the proportions in instance $I_p$ in round $t$, induced by training scheme $\bm x$. Furthermore, we denote $V_T^p(\bm x)$ the revenue over $T$ rounds in instance $I_p$, induced by $\bm x$. We use the following lemma to establish the relationship between $T(p)$ and $T(p')$ for $p > p'$.

From Lemma~\ref{prop greater p0}  we get that for every $T$ and every $p > p'$ it holds that $V_T^p(\bm x) \geq V_T^{p'}(\bm x)$. Recall that $T^p$ is the minimal number of rounds it takes for $V_{T^p}^p(\bm x^0) < 0$ and therefore $T^p > T^{p'}$.
Lastly, Notice that $T_0 = T^1 = \max_p T^p$.

Next, we use the following lemma to show that for a given instance $I_p$, if there is a training scheme that induces a non-negative revenue for any $T > T^p$, then we can bound the time of the first training round of the optimal training scheme by the first horizon $T$ to satisfy $V_T^p(\bm x) \geq 0$.

\begin{lemma} \label{bounded first training step}
Given any $p \in [0,1]$, if there exists $\bm x$ and $T' \geq T^p$ such that $V_{T'}^p \geq 0$ then for every $T > T'$ there exists $t \leq T'$ such that optimal training scheme in $I^p$ satisfies $x^\star_t = 1$.
\end{lemma}

Therefore, if there exists a scheme $\bm x'$ that satisfies $V_{T}^0(\bm x') \geq 0$ for $T' > T_0$ then by Lemma~\ref{prop greater p0}, for every $p \in [0, 1]$ it holds that $V_{T'}^p(\bm x') \geq 0$. Since $T \geq T_0 \geq T^p$ then by Lemma~\ref{bounded first training step}, there exists $t \leq T'$ such that for every $T \geq T'$, the optimal training scheme satisfies $\bm x^\star = 1$ on instance $I_p$.

Lastly, Notice that the optimal training scheme is optimal for any subproblem which starts at a training round $t \in [0, T]$ with horizon $T-t$. Therefore, it is also the optimal training scheme for the subproblem starting from the training round $\tau^\star_i$ with proportion $p_{\tau^\star_i}(\bm x^\star)$ and horizon $T - \tau^\star_i$. Thus, if $T-\tau^\star_i \geq T'$ then the optimal training scheme has a training round in $t \in (\tau^\star_i, \tau^\star_i + T']$.
\end{proofof}

\begin{proofof}{bounded first training step}
Notice that for the instance with horizon $T \geq T' \geq T^p$ it holds that $V(\bm x^0) < 0$ and $V(\bm x) \geq 0 > V(\bm x^0)$. Therefore, not training is not optimal. Thus, the optimal training scheme $\bm x^\star$ must have at least one training round besides $t = 1$. 

We define \emph{horizon-dependent instance} with horizon $T$ as $J = (\rvg, c_m, c_{train}, \rg, \ro, \beta, p, T)$ and analyze two instances with horizon $T'$ and $T > T'$ respectively:
\begin{align*}
    & J_1 = (\rg, \ro, \beta, p, \rvg, c_m, c_{train}, T'), \\
    & J_2 = (\rg, \ro, \beta, p, \rvg, c_m, c_{train}, T).
\end{align*}

Let $\tau > 1$ be the first training round of $\bm x$ in instance $J_1$, and let $\tau^\star > 1$ be the first training round of the optimal training scheme $\bm x^\star$ in instance $J_2$. We show that there is no $T > T'$ such that $\tau^\star > T'$.

Assume in contradiction that there exists $T$ and $J_2$ such that $\tau^\star > T'$, i.e., the optimal training scheme $\bm x^\star$ does not train before round $t = T'$. 
We define the training scheme $\bm x'$ on instance $J_2$ as
\begin{align*}
x'_t = \begin{cases}
    x_t & \mbox{$t \leq T'$} \\
    x^\star_t & \mbox{$t > T'$}
\end{cases}.
\end{align*}
Notice that
\begin{align*}
& x'_t = x_t^\star = x_t \qquad \forall t < \tau \\
& x^\star_t = x^0_t      \qquad \qquad \forall t < \tau^\star,
\end{align*}
therefore,
\begin{align*}
T \left( V(\bm x') - V(\bm x^\star) \right) &= \sum_{t = 1}^{T} v_t(\bm x') - v_t(\bm x^\star) \\
&= \sum_{t = 1}^{\tau} \left( v_t(\bm x') - v_t(\bm x^\star) \right) + \sum_{t = \tau + 1}^{T'} \left( v_t(\bm x') - v_t(\bm x^\star) \right) + \sum_{t = T' + 1}^{T} \left( v_t(\bm x') - v_t(\bm x^\star) \right) \\
&= \sum_{t = 1}^{\tau} \left( v_t(\bm x^\star) - v_t(\bm x^\star) \right) + \sum_{t = \tau + 1}^{T'} \left( v_t(\bm x) - v_t(\bm x^0) \right) + \sum_{t = T' + 1}^{T} \left( v_t(\bm x') - v_t(\bm x^\star) \right) \\
&= \sum_{t = \tau + 1}^{T'} \left( v_t(\bm x) - v_t(\bm x^0) \right) + \sum_{t = T' + 1}^{T} \left( v_t(\bm x') - v_t(\bm x^\star) \right).
\end{align*}
Using the above, we have
\begin{align} \label{eq j2}
T \left( V(\bm x') - V(\bm x^\star) \right) = \sum_{t = \tau + 1}^{T'} \left( v_t(\bm x) - v_t(\bm x^0) \right) + \sum_{t = T' + 1}^{T} \left( v_t(\bm x') - v_t(\bm x^\star) \right),
\end{align}
but on $J_1$ we got $V(\bm x) \geq 0 > V(\bm x^0)$ and therefore
\begin{align*}
0 &< \sum_{t = 1}^{T'} v_t(\bm x) - v_t(\bm x^0) \\
& = \sum_{t = 1}^{\tau} \left( v_t(\bm x) - v_t(\bm x^0) \right) + \sum_{t = \tau + 1}^{T'} \left( v_t(\bm x) - v_t(\bm x^0) \right) \\
& = \sum_{t = 1}^{\tau} \left( v_t(\bm x^0) - v_t(\bm x^0) \right) + \sum_{t = \tau + 1}^{T'} \left( v_t(\bm x) - v_t(\bm x^0) \right) \\
& = \sum_{t = \tau + 1}^{T'} \left( v_t(\bm x) - v_t(\bm x^0) \right).
\end{align*}
Overall, the first summation in Equation~\ref{eq j2} is positive. The second summation is also positive from Lemma~\ref{prop greater p0}, which results in $V(\bm x') - V(\bm x) > 0$. In other words, scheme $\bm x^\star$ is not optimal in contradiction to our assumption.
\end{proofof}

\begin{proofof}{close probabilities contracting}
Consider two new instance with horizon $T = k$, namely $I_1$ and $I_2$ where the initial proportion of $I_1$ and $I_2$ are $p^1$ and $p^2$ respectively. Further, consider the training scheme $\bm x^0$ that does not train after $t = 1$.
Therefore, this lemma is a special case of Theorem~\ref{thm: contracting}.

Thus, according to the proof of Theorem~\ref{thm: contracting}, $\gamma$ is defined by
\[
\gamma = \frac{1}{1 + 2 e^{\beta L} \delta },
\]
where $\delta = \frac{16-7\beta L}{14 \beta L e^{\beta L}} - 0.002 = 0.234$. Hence, $\gamma = 0.44$.
\end{proofof}

\begin{proofof}{opt near vicinity}
Let $\varepsilon = \left| f(p) - p \right|$. We use the following lemma.

\begin{lemma} \label{distance from next}
For every $n \in \mathbb{N}$ it holds that $p - \sum_{i = 0}^{n-1} \gamma^i \varepsilon \leq f^{(n)}(p) \leq p + \sum_{i = 0}^{n-1} \gamma^i \varepsilon$.
\end{lemma}
By definition $p^\star = \lim_{n \rightarrow \infty} f^{n}(p)$. Therefore, according to Lemma~\ref{distance from next}
\[
p - \varepsilon \sum_{i = 0}^\infty \gamma^i \leq p^\star \leq p + \varepsilon \sum_{i = 0}^\infty \gamma^i.
\]
The term $\varepsilon \sum_{i = 0}^\infty \gamma^i$ is the sum of a geometric series with factor $\gamma < 1$ and therefore $\varepsilon \sum_{i = 0}^\infty \gamma^i = \frac{\varepsilon}{1-\gamma}$. Thus
\[
p - \frac{\varepsilon}{1-\gamma} \leq p^\star \leq p + \frac{\varepsilon}{1-\gamma},
\]
which implies
$\left|p - p^\star \right| \leq \frac{\varepsilon}{1-\gamma}$.
This concludes the proof.
\end{proofof}

\begin{proofof}{distance from next}
We prove this by induction. Starting with $n = 1$, the equality $\left|f(p) - p \right| = \varepsilon$ implies $f(p) - p = \varepsilon$ or $f(p) - p = -\varepsilon$. In the first case $f(p) = \varepsilon + p$. In the second case, $f(p) = -\varepsilon + p$. Next, we assume it is true for every $n$, and we prove it for $n+1$. Therefore, we start with inequality: 
\[
\left| f^{(n+1)}(p) - f^{(n)}(p)\right| \leq \gamma \left| f^{(n)}(p) - f^{(n-1)}(p) \right| \leq \gamma^{n} \varepsilon
\]
Which implies 
\[
-\gamma^{n} \varepsilon + f^{(n)}(p) \leq f^{(n+1)}(p) \leq \gamma^{n} \varepsilon + f^{(n)}(p).\] 
We use the assumption regarding $f^{(n)}(p)$ and get
\[
p - \sum_{i = 0}^{n}\gamma^i \varepsilon \leq f^{(n+1)}(p) \leq p + \sum_{i = 0}^{n}\gamma^i \varepsilon.
\]
\end{proofof}

\begin{proofof}{obs: general a is recursive}
This is a special case of Observation~\ref{obs: transition from pt0} with $t_0 = t_1 = 0$ and $k = \alpha_1 + \alpha_2$.
\end{proofof}
\section{Proofs Omitted from Section~\ref{section social impact}}

\begin{proofof}{obs possible harmful}
Consider the instance $\rvg = 1$, $c_m = 0.1$, $\rg(t) = 3\cdot 0.5^t + 0.5$, $\ro(p) = 0.51-p$, $\beta = 1$, $p_1 = 1$. We set $c_{train} = T$ and show how to pick $T$ by the end of this proof. Let $\bm x^0$ be the training scheme that trains only at $t = 1$, i.e $x_1 = 1$ and $x_t = 0$ for every $t > 1$.

$c_{train}$ is chosen such that $\bm x^0$ is optimal. Therefore we aim for the change in the proportion in each round until $T$ to induce less revenue then $c_{train}$

Let $p^- = \frac{e^{0.5}}{e^{0.5} + e^{0.51}} = 0.497$ and notice that for every $p \in [0, 1]$ and every $t \in [T]$ it holds 
\begin{align*}
p^- = \frac{e^{0.5}}{e^{0.5} + e^{0.51}} = \frac{e^{\beta \inf_{t'} \rg(t')}}{e^{\beta \inf_{t'} \rg(t')} + e^{\ro(1)}} \leq \frac{e^{\beta \inf_{t'} \rg(t')}}{e^{\beta \inf_{t'} \rg(t')} + e^{\ro(p)}} \leq \frac{e^{\beta \rg(t)}}{e^{\beta \rg(t)} + e^{\ro(p)}}.
\end{align*}
Therefore for any $t \in [T]$ it holds $p_t(\bm x^0) \geq p^-$. The revenue of \llm at time $t$ is 
\[
v_t(\bm x^0) = p_t(\bm x^0)\rvg - c_{train} \geq 0.497\cdot 1 - 0.1 = 0.397. 
\]

for every $t \in [T]$ it holds $1 \geq v_t \geq 0.397$ and therefore $T\cdot V(\bm x^0) \leq T$. In other words, if \llm were to train in any round $t > 1$ then the induced revenue would have been negative. Thus, $\bm x^0$ is optimal.

Let $t_0$ be the round for which $\rg(t_0 - 1) < \ro(0)$, and notice that for every $t \geq t_0$ it holds that
\begin{align*}
u_t(\bm x^0) &= p_t(\bm x^0)\rg(t - 1) + (1-p_t(\bm x^0)) \ro(p_t(\bm x^0)) \\
&\leq p_t(\bm x^0)\rg(t_0 - 1) + (1-p_t(\bm x^0)) \ro(p_t(\bm x^0)) \\
&\leq p^-\rg(t_0 - 1) + (1-p^-) \ro(p_t(\bm x^0)) \leq p^- \rg(t_0 - 1) + (1-p^-)\ro(p^-) < \ro(0).
\end{align*}

let
\[
\varepsilon = \ro(0) - p^- \rg(t_0 - 1) + (1-p^-)\ro(p^-).
\]

Thus, The difference in the cumulative welfare with and without \llm is
\begin{align*}
T\ro(0) - U(\bm x^0) = T\ro(0) - \sum_{t = 1}^{T}u_t(\bm x^0) = T\ro(0) - \sum_{t = 1}^{t_0 - 1}u_t - \sum_{t = t_1}^T u_t(\bm x^0) \geq (T-t_0)\ro(0) - \sum_{t = 1}^{t_0 - 1}u_t + (T-t_0)\varepsilon.
\end{align*}

Therefore for $T > \frac{\sum_{t = 1}^{t_0 - 1}u_t - (T-t_0)\ro(0) + t_0\varepsilon}{\varepsilon}$ it holds $T\ro(0) - U(\bm x^0) > 0$.
\end{proofof}

\subsubsection{Proofs Omitted from Subsection~\ref{subsec: social PoA}}
\begin{proofof}{infinite poa}
Given $T \in \mathbb{R}$, we explain how we construct our instance. We choose the value of $T$ by the end of the proof. Consider the instance with the following parameters: $\rvg = 1$, $c_m = 0.6, c_{train} = 2T$, $\rg(t) = 3 \cdot 0.5^t$, $\ro(p) = 1-p$, $\beta = \infty$, $p_1 = 1$.

We first highlight a property of our instance, which is a product of $\beta = \infty$. Given a training scheme $\bm x$, for every $t \in [T]$, $p_t(\bm x)$ satisfies

\begin{align*}
p_{t+1}(\bm{x}) = \begin{cases}
    1 & \mbox{$\rg(\gamma_t) > \ro(p_t(\bm x))$} \\
    0.5 & \mbox{$\rg(\gamma_t) = \ro(p_t(\bm x))$} \\
    0 & \mbox{Otherwise}
\end{cases}.
\end{align*}

This property leads to the following observation.
\begin{observation} \label{obs: best response prop}
For every training scheme $\bm x$ and round $t \in [T]$ it holds $p_t(\bm x) = 1$.
\end{observation}

We first show that the welfare-maximizing training scheme trains in every round. Observe the instantaneous welfare at round $t$
\[
u_t = \rg(\gamma_t) \leq \rg(0),
\]
where the equality holds only for $\gamma_t$, and therefore \llm has to train in every round to maximize the social welfare. Thus
\[
\max_{\bm x}U(\bm x) = T\rg(0).
\]

Next, we examine \llm related attributes. An immediate conclusion from Observation~\ref{obs: best response prop} is that $p_t(\bm x) = 1$ is constant and does not depend on the scheme $\bm x$. In other words, the revenue of \llm is
\[
V = \frac{T\rvg - Tc_m - c_{train}\sum_{t = 1}^T \ind_{x_t = 1}}{T} = 0.4 - \frac{c_{train}}{T} \sum_{t = 1}^T \ind_{x_t = 1}.
\]

Therefore, the optimal training scheme that maximizes the revenue is the training scheme that trains the least amount of times, i.e., $x_t = 0$ for every $t > 1$. Thus, the social welfare under the revenue-maximizing scheme is
\[
\min_{\bm x \in \mathcal{X}}U(\bm x) = \sum_{t = 1}^T \rg(t - 1).
\]

Let $\varepsilon > 0$ then according to the decreasing assumption on $\rg(t)$, and there exists $t_0$ such that for every $t > t_0$ it holds $\rg(t) < \varepsilon$. Therefore, for $T > t_0$, the price of anarchy is
\begin{align*}
\poa(I) = \frac{T\rg(0)}{\sum_{t = 1}^{t_0} \rg(t-1) + \varepsilon(T-t_0)}.
\end{align*}

There exists $T_0$ such that for every $T > T_0$ it holds $\varepsilon(T-t_0) > \sum_{t = 1}^{t_0} \rg(t)$, thus
\begin{align*}
\poa(I) > \frac{T\rg(0)}{2(T-t_0)\varepsilon} > \frac{T\rg(0)}{2T\varepsilon} = \frac{\rg(0)}{2\varepsilon}.
\end{align*}

Thus, $\poa(I) > M$ for $\varepsilon = \frac{\rg(0)}{2M}$ and $T >  \frac{t_0 \rg(0) + 2M\sum_{t = 1}^{t_0} \rg(t-1)}{\rg(0)}$.

\end{proofof}

\begin{proofof}{obs: best response prop}
Fix any arbitrary training scheme, and we prove by induction over $t$. The base case for $t = 1$ trivially holds. Assume it is true for $t-1$ and we prove our induction for $t$.
Notice that $\ro(p_{t-1}(\bm x)) = \ro(1) = 0$ and $\rg(t) > 0 = \ro(p_{t-1}(\bm x))$, therefore from the property of $\beta = \infty$ it holds $p_t(\bm x) = 1$.
\end{proofof}
\section{Proofs Omitted from Section~\ref{subsec:regulator}}

\subsection{Proofs Omitted from Subsection~\ref{subsec: bounding proportions}}

\begin{proofof}{prop greater p0}
We show it by induction over $t$. The base case trivially holds from the condition of the lemma. Next we assume the lemma holds for $t-1$ and prove it for $t$. Notice that $p'_{t-1}(\bm x) > p_{t-1}(\bm x)$ results in $\ro(p'_{t-1}(\bm x)) < \ro(p_{t-1}(\bm x))$. Therefore,

\begin{align*}
p'_t(\bm x)  = \frac{e^{\beta \rg(\gamma_t)}}{e^{\beta \rg(\gamma_t)} + e^{\beta \ro(p'_{t-1}(\bm x))}} > \frac{e^{\beta \rg(\gamma_t)}}{e^{\beta \rg(\gamma_t)} + e^{\beta \ro(p_{t-1}(\bm x))}} = p_t(\bm x).
\end{align*}
\end{proofof}

\begin{proofof}{thm: sufficient helpful}
We start with the sufficient condition by showing that for every $t \in [0, \Delta]$, it holds that
\[
q^0_t \rg(t) + (1-q^1_t)\ro(q^1_t) \leq u_{\tau + t}(\bm x).
\]

Therefore, by Inequality~\eqref{eq:sandwich p} we get
\begin{align*}
u_{\tau + t}(\bm x) &= p_t(\bm x) \rg(\gamma_t) + (1-p_t(\bm x)) \ro(p_t(\bm x)) \\
& \geq q^0_t \rg(\gamma_t) + (1-p_t(\bm x)) \ro(p_t(\bm x)) \\
& \geq q^0_t \rg(\gamma_t) + (1-q^1_t) \ro(q^1_t) = \unu_t.
\end{align*}

Thus, we can also bound the summation. Hence, sufficient condition leads to the following:
\begin{align*}
\Delta \ro(0) \leq \sum_{t = 0}^{\Delta - 1} \unu_t \leq \sum_{t = 0}^{\Delta - 1} u_t(\bm x).
\end{align*}

In the same manner, we prove the necessary condition. First we see that
\begin{align*}
u_{\tau + t}(\bm x) &= p_t(\bm x) \rg(\gamma_t) + (1-p_t(\bm x)) \ro(p_t(\bm x)) \\
&\leq  q^1_t \rg(\gamma_t) + (1-p_t(\bm x)) \ro(p_t(\bm x)) \\
&\leq q^1_t \rg(\gamma_t) + (1-q^0_t) \ro(q^0_t) = \ovu_t.
\end{align*}

Therefore, if \llm is socially beneficial, then it holds that
\begin{align*}
\Delta \ro(0) \leq \sum_{t = 0}^{\Delta - 1} u_t(\bm x) \leq \sum_{t = 0}^{\Delta - 1} \ovu_t.
\end{align*}

\end{proofof}

\subsection{Proofs Omitted from Subsection~\ref{subsec:noisy}}

\begin{proofof}{thm: contracting}
Recall the definition of transition functions from Section~\ref{sec: transition functions}. We show that this is a property of transition functions. Observe that
\[
q_{t+1}^\alpha = f_t(q_t^{\alpha}).
\]
Therefore, to prove this theorem, we aim to prove a contraction property of $f_t$. We start by finding the relationship between $\left| f_t(p_{\tau + t}(\bm x)) - f_t(q_t^\alpha) \right|$ and $\left| p_{\tau + t}(\bm x) - q_t^\alpha \right|$ using the following lemma

\begin{lemma} \label{transition contracting equation}
Let $f_t(p) = \frac{e^{\beta \rg(t)}}{e^{\beta \rg(t)} + e^{\beta \ro(p)}}$ and let $p^1, p^2 \in [0, 1]$. If $\ro$ is $L$-Lipschitz with $L \leq 1$ and $|p^1 - p^2| < \frac{1}{\beta L}$ then
\[
\left| f_t(p^1) - f_t(p^2) \right| < \frac{1}{4 - \frac{7}{2} \beta L e^{\beta \left( \ro(p^1) - \rg(t) \right)} \left| p^2 - p^1 \right|} \frac{7}{4} \beta L \abs{p^1 - p^2}.
\]
\end{lemma}

Notice that since $\beta L < \frac{16}{7}$ it holds $\frac{16-7\beta L}{14 \beta L e^{\beta L}} < \frac{1}{\beta L}$. Therefore, any $\varepsilon$ that satisfies this theorem ensures that $\abs{p_{\tau}(\bm x) - q^\alpha_0}$ also satisfies the conditions of Lemma~\ref{transition contracting equation}. Thus, by Lemma~\ref{transition contracting equation} we have

\[
\left| f_t(p_{\tau + t}(\bm x)) - f_t(q_t^\alpha) \right| < \frac{1}{4 - \frac{7}{2} \beta L e^{\beta \left( \ro(p_{\tau + t}(\bm x)) - \rg(t) \right)} \left| p_{\tau + t}(\bm x) - q_t^\alpha \right|} \frac{7}{4} \beta L \left| p_{\tau + t}(\bm x) - q_t^\alpha \right|.
\]

Denote $\delta = \frac{16-7\beta L}{14 \beta L e^{\beta L}} - \varepsilon$ then for $t = 0$ it holds
\begin{align*}
\left| f_0(p_{\tau}(\bm x)) - f_0(q_0^{\alpha}) \right| &< \frac{1}{4 - \frac{7}{2} \beta L e^{\beta \left( \ro(p_{\tau}(\bm x)) - \rg(t) \right)} \left(\frac{16-7\beta L}{14 \beta L e^{\beta L}} - \delta \right)} \frac{7}{4} \beta L \abs{p_{\tau}(\bm x) - q_0^{\alpha}} \\
&\leq \frac{1}{4 - \frac{7}{2} \beta L e^{\beta L} \left(\frac{16-7\beta L}{14 \beta L e^{\beta L}} - \delta \right)} \frac{7}{4} \beta L \abs{p_{\tau}(\bm x) - q_0^{\alpha}} \\
&= \frac{1}{1 + 2 e^{\beta L} \delta }  \abs{p_{\tau}(\bm x) - q_0^{\alpha}}.
\end{align*}

Denote $\gamma = \frac{1}{1 + 2 e^{\beta L} \delta }$ and notice that $\gamma < 1$.
Next, we show that $\abs{p_{\tau+t}(\bm x) - q^\alpha_t} < \varepsilon \cdot \gamma^t$ by induction. The theorem's condition trivially satisfies the base case for $t = 0$. Assume the theorem holds for $t$, and we prove it for $t+1$. Then,

\begin{align*}
\abs{p_{\tau+t+1}(\bm x) - q^\alpha_{t + 1}} &= \abs{ f_t(p_{\tau}(\bm x)) - f_t(q^\alpha_t) } \\
&< \frac{1}{4 - \frac{7}{2} \beta L e^{\beta \left( \ro(p_{t}(\bm x)) - \rg(t) \right)}} \abs{p_{t}(\bm x) - q_{t}^{\alpha}} \frac{7}{4} \beta L \abs{p_{t}(\bm x) - q_{t}^{\alpha}} \\
& \leq \frac{1}{4 - \frac{7}{2} \beta L e^{\beta L} \abs{p_{t}(\bm x) - q_{t}^{\alpha}}} \frac{7}{4} \beta L \abs{p_{t}(\bm x) - q_{t}^{\alpha}} \\
& \leq \frac{1}{4 - \frac{7}{2} \beta L e^{\beta L} \gamma^{t}\abs{p_{\tau}(\bm x) - q_0^{\alpha}}} \frac{7}{4} \beta L \abs{p_{t}(\bm x) - q_{t}^{\alpha}} \\
& \leq \frac{1}{4 - \frac{7}{2} \beta L e^{\beta L} \abs{p_{\tau}(\bm x) - q_0^{\alpha}}} \frac{7}{4} \beta L \abs{p_{t}(\bm x) - q_{t}^{\alpha}} \\
& \leq \gamma \abs{p_{t}(\bm x) - q_{t}^{\alpha}} \\
& \leq \gamma \gamma^{t} \abs{p_{\tau}(\bm x) - q_0^{\alpha}} \\
&= \gamma^{t+1} \abs{p_{\tau}(\bm x) - q_0^{\alpha}} < \gamma^{t+1} \varepsilon.
\end{align*}
\end{proofof}

\begin{proofof}{transition contracting equation}
We start by showing the relationship between $\abs{f_t(p^1) - f_t(p^2)}$ and $\abs{p^1 - p^2}$.

By definition:
\begin{align*}
\abs{f_t(p^1) - f_t(p^2)} &= \left| \frac{e^{\beta \rg(t)}}{e^{\beta \rg(t)} + e^{\beta \ro(p^1)}} - \frac{e^{\beta \rg(t)}}{e^{\beta \rg(t)} + e^{\beta \ro(p^2)}} \right| \\
&= \left| e^{\beta \rg(t)} \frac{e^{\beta \ro(p^2)} - e^{\beta \ro(p^1)}}{\left( e^{\beta \rg(t)} + e^{\beta \ro(p^1)} \right)\left( e^{\beta \rg(t)} + e^{\beta \ro(p^2)} \right)}  \right| \\
&= \left|f_t(p^1) \frac{e^{\beta \ro(p^2)} - e^{\beta \ro(p^1)}}{e^{\beta \rg(t)} + e^{\beta \ro(p^2)}} \right| \\
&= \left|f_t(p^1) \frac{1 - e^{\beta \left(\ro(p^1) - \ro(p^2)\right)}}{e^{\beta \left(\rg(t) - \ro(p^2)\right)} + 1} \right| \\
&= \left|f_t(p^1) (1-f_t(p^2)) \left( 1 - e^{\beta \left(\ro(p^1) - \ro(p^2)\right)} \right) \right| \\
&= f_t(p^1) (1-f_t(p^2)) \left| 1 - e^{\beta \left(\ro(p^1) - \ro(p^2)\right)} \right|.
\end{align*}

Since $\ro$ is $L$-Lipschitz, and $\abs{p^1 - p^2} < \frac{1}{\beta L}$, it holds 
\[
\left| \beta \left(\ro(p^1) - \ro(p^2)\right) \right| \leq \beta L \abs{p^1 - p^2} < 1.
\]
Therefore, we use the inequality $\abs{e^x - 1} < \frac{7}{4}\abs{x}$ for $\abs{x} < 1$ and get the following:
\begin{align*}
\abs{f_t(p^1) - f_t(p^2)}  &< f_t(p^1) (1-f_t(p^2)) \frac{7}{4} \beta \left| \ro(p^1) - \ro(p^2) \right| \\
&\leq f_t(p^1) (1-f_t(p^2)) \frac{7}{4} \beta L \left| p^1 - p^2 \right|.
\end{align*}

We expand the expression for $f_t(p^2)$. Notice that

\begin{align*}
\frac{e^{\beta \rg(t)}}{e^{\beta \rg(t)} + e^{\beta \ro(p^2)}} &= \frac{e^{\beta \rg(t)}}{e^{\beta \rg(t)} + e^{\beta \ro(p^1)} - e^{\beta \ro(p^1)} + e^{\beta \ro(p^2)}} \\
&= \frac{e^{\beta \rg(t)}}{e^{\beta \rg(t)} + e^{\beta \ro(p^1)} + e^{\beta \ro(p^1)}\left( e^{\beta \left( \ro(p^2) - \ro(p^1)\right)} - 1 \right)} \\
&\geq \frac{e^{\beta \rg(t)}}{e^{\beta \rg(t)} + e^{\beta \ro(p^1)} + e^{\beta \ro(p^1)}\left| e^{\beta \left( \ro(p^2) - \ro(p^1)\right)} - 1 \right|} \\
&\geq \frac{e^{\beta \rg(t)}}{e^{\beta \rg(t)} + e^{\beta \ro(p^1)} + e^{\beta \ro(p^1)} \frac{7}{4} \left| \beta \left( \ro(p^2) - \ro(p^1)\right) \right|} \\
&\geq \frac{e^{\beta \rg(t)}}{e^{\beta \rg(t)} + e^{\beta \ro(p^1)} + e^{\beta \ro(p^1)} \frac{7}{4} \beta L \left| p^2 - p^1 \right|} \\
&= \left(\frac{1}{f_t(p^1)} + \frac{7}{4} \beta L e^{\beta \left( \ro(p^1) - \rg(t) \right)} \left| p^2 - p^1 \right| \right)^{-1}.
\end{align*}

Therefore, we can write:
\begin{align*}
f_t(p^1) (1 - f_t(p^2)) &\leq f_t(p^1) \left(1 - \left(\frac{1}{f_t(p^1)} + \frac{7}{4} \beta L e^{\beta \left( \ro(p^1) - \rg(t) \right)} \left| p^2 - p^1 \right| \right)^{-1} \right)
\end{align*}

and finally
\begin{align*}
\abs{f_t(p^1) - f_t(p^2)} < f_t(p^1) \left(1 - \left(\frac{1}{f_t(p^1)} + \frac{7}{4} \beta L e^{\beta \left( \ro(p^1) - \rg(t) \right)} \left| p^2 - p^1 \right| \right)^{-1} \right) \frac{7}{4} \beta L \left| p^1 - p^2 \right|.
\end{align*}

The expression $x\left(1-\left(\frac{1}{x} +y\right)^{-1}\right)$ has one maximum point at $x^\star = \frac{1}{2-y}$, therefore by setting $y = \frac{7}{4} \beta L e^{\beta \left( \ro(p^1) - \rg(t) \right)} \left| p^2 - p^1 \right|$ and $x = f_t(p^1)$ we get
\[
x^\star = \frac{1}{2 - \frac{7}{4} \beta L e^{\beta \left( \ro(p^1) - \rg(t) \right)} \left| p^2 - p^1 \right|}.
\]

Plugging in the result
\begin{align*}
&f_t(p^1) \left(1 - \left(\frac{1}{f_t(p^1)} + \frac{7}{4} \beta L e^{\beta \left( \ro(p^1) - \rg(t) \right)} \left| p^2 - p^1 \right| \right)^{-1} \right) \\
&\leq \frac{1}{4 - \frac{7}{2} \beta L e^{\beta \left( \ro(p^1) - \rg(t) \right)} \left| p^2 - p^1 \right|}.    
\end{align*}

Thus, we finally get
\[
\left| f_t(p^1) - f_t(p^2) \right| < \frac{1}{4 - \frac{7}{2} \beta L e^{\beta \left( \ro(p^1) - \rg(t) \right)} \left| p^2 - p^1 \right|} \frac{7}{4} \beta L |p^1 - p^2|.
\]
\end{proofof}

\begin{proofof}{prop:tight bound on u}
First, notice that by Corollary~\ref{cor: bounds with q} and definitions of $\ovq_t$, $\unq_t$ we get for every $t \in [0, \Delta-1]$
\begin{align*}
& 0 \leq p_{\tau + t}(\bm x) - \unq_t \leq 2\varepsilon \gamma^t,  \\
& 0 \leq \ovq_t - p_{\tau + t}(\bm x) \leq 2\varepsilon \gamma^t. 
\end{align*}

We start with the first inequality. For every $t \in [0, \Delta - 1]$ it holds that
\begin{align*}
u_{\tau + t}(\bm x) - \unu_t^\varepsilon &= p_{\tau + t}(\bm x) \rg(t) + (1-p_{\tau + t}(\bm x))\ro(p_{\tau + t}(\bm x)) - \unq_t \rg(t) - (1-\ovq_t)\ro(\ovq_t) \\
&= ( p_{\tau + t}(\bm x) - \unq_t ) \rg(t) + (1-p_{\tau + t}(\bm x))\ro(p_{\tau + t}(\bm x)) - (1-\ovq_t)\ro(\ovq_t) \\
&\leq 2\varepsilon \gamma^t \rg(t) + (1-p_{\tau + t}(\bm x))\ro(p_{\tau + t}(\bm x)) - (1-\ovq_t)\ro(\ovq_t) \\
&= 2\varepsilon \gamma^t \rg(t) + (1-p_{\tau + t}(\bm x))\ro(p_{\tau + t}(\bm x)) - (1-\ovq_t + p_{\tau + t}(\bm x) - p_{\tau + t}(\bm x))\ro(\ovq_t) \\
&= 2\varepsilon \gamma^t \rg(t) + (1-p_{\tau + t}) \left( \ro(p_{\tau + t}(\bm x)) - \ro(\ovq_t) \right) + \left(\ovq_t - p_{\tau + t}(\bm x) \right) \ro(\ovq_t) \\
&\leq 2\varepsilon \gamma^t \rg(t) + \ro(p_{\tau + t}(\bm x)) - \ro(\ovq_t) +  \left( \ovq_t - p_{\tau + t}(\bm x) \right) \ro(\ovq_t) \\
&\leq 2\varepsilon \gamma^t \rg(t) + 2L\varepsilon \gamma^t + \left(\ovq_t - p_{\tau + t}(\bm x) \right) \ro(\ovq_t) \\
&\leq 2\varepsilon \gamma^t \rg(t) + 2L\varepsilon \gamma^t + 2\varepsilon \gamma^t \ro(\ovq_t) \\
&\leq 2\varepsilon \gamma^t \rg(t) + 2L\varepsilon \gamma^t + 2\varepsilon L \gamma^t \\
&= 2\varepsilon \gamma^t \left(\rg(t) + 2L\right).
\end{align*}

We now move on to the second condition.
\begin{align*}
\ovu_t^\varepsilon - u_{\tau + t}(\bm x) &= \ovq_t \rg(t) + (1-\unq_t)\ro(\unq_t) - p_t(\bm x)\rg(t) - (1-p_t(\bm x))\ro(p_t(\bm x)) \\
&= \left( \ovq_t - p_t(\bm x) \right)\rg(t) + (1-\unq_t)\ro(\unq_t) - (1-p_t(\bm x))\ro(p_t(\bm x)) \\
&\leq 2\varepsilon \gamma^t \rg(t) + (1-\unq_t)\ro(\unq_t) - (1-p_t(\bm x))\ro(p_t(\bm x)) \\
&= 2\varepsilon \gamma^t \rg(t) + ( 1-\unq_t + p_t(\bm x) - p_t(\bm x) ) \ro(\unq_t) - (1-p_t(\bm x))\ro(p_t(\bm x)) \\
&= 2\varepsilon \gamma^t \rg(t) + (1 - p_t(\bm x))(\ro(\unq_t) - \ro(p_t(\bm x))) + (p_t(\bm x) -\unq_t) \ro(\unq_t) \\
&\leq 2\varepsilon \gamma^t \rg(t) + (\ro(\unq_t) - \ro(p_t(\bm x))) + (p_t(\bm x) -\unq_t) \ro(\unq_t) \\
&\leq 2\varepsilon \gamma^t \rg(t) + 2L\varepsilon \gamma^t + (p_t(\bm x) -\unq_t) \ro(\unq_t) \\
&\leq 2\varepsilon \gamma^t \rg(t) + 2L\varepsilon \gamma^t + 2L\varepsilon \gamma^t \\
&= 2\varepsilon \gamma^t \left(\rg(t) + 2L \right).
\end{align*}
\end{proofof}

\begin{proofof}{thm:u star is not far}
From Proposition~\ref{prop:tight bound on u}:

\begin{align*}
\sum_{t = 0}^{\Delta - 1}\ovu^{\varepsilon}_{t}  - \sum_{t = 0}^{\Delta - 1}\unu^{\varepsilon}_{t} &= \sum_{t = 0}^{\Delta - 1}\ovu^{\varepsilon}_{t} - \sum_{t = 0}^{\Delta - 1} u_t(\bm x) + \sum_{t = 0}^{\Delta - 1} u_t(\bm x) - \sum_{t = 0}^{\Delta - 1}\unu^{\varepsilon}_{t} \\
&= \sum_{t = 0}^{\Delta - 1}\left(\ovu^{\varepsilon}_{t} - u_t(\bm x)\right) + \sum_{t = 0}^{\Delta - 1} \left(u_t(\bm x) - \unu^{\varepsilon}_{t}\right) \\
&\leq \sum_{t = 0}^{\Delta - 1} \left( 2\varepsilon \gamma^t (\rg(t) + 2L) \right) + \sum_{t = 0}^{\Delta - 1} \left( 2\varepsilon \gamma^t (\rg(t) + 2L) \right) \\
&= 4\varepsilon \sum_{t = 0}^{\Delta - 1} \gamma^t (\rg(t) + 2L).
\end{align*}
\end{proofof}
}{\fi}

\end{document}